\documentclass{aastex631}
\usepackage{xcolor}
\usepackage{tabu}

\newcommand{\CC}[1]{}

\usepackage{rotating}

\shorttitle{M~dwarf Exoplanet Hosts}
\shortauthors{Matson et al.}
\graphicspath{{./}{figures/}}

\begin{document}

\title{Demographics of M~Dwarf Binary Exoplanet Hosts Discovered by TESS}

\correspondingauthor{Rachel Matson}
\email{rachel.a.matson2.civ@us.navy.mil}

\author[0000-0001-7233-7508]{Rachel A. Matson}
\affiliation{U.S. Naval Observatory, 3450 Massachusetts Avenue NW, Washington, D.C. 20392, USA}

\author[0009-0000-7274-7523]{{Rebecca Gore}}
\affiliation{Bay Area Environmental Research Institute, Moffett Field, CA 94035, USA}

\author[0000-0002-2532-2853]{Steve~B.~Howell}
\affil{NASA Ames Research Center, Moffett Field, CA 94035, USA}

\author[0000-0002-5741-3047]{David R.~Ciardi}
\affiliation{NASA Exoplanet Science Institute, Caltech/IPAC, 1200 E. California Blvd., Pasadena, CA 91125, USA}

\author[0000-0002-8035-4778]{Jessie L.\ Christiansen}
\affiliation{NASA Exoplanet Science Institute, Caltech/IPAC, 1200 E. California Blvd., Pasadena, CA 91125, USA}

\author[0000-0002-2361-5812]{Catherine A.~Clark}
\affiliation{NASA Exoplanet Science Institute, Caltech/IPAC, 1200 E. California Blvd., Pasadena, CA 91125, USA}

\author{Ian J.\ M.\ Crossfield}
\affil{Department of Physics and Astronomy, University of Kansas, Lawrence, KS 66045, USA}

\author[0000-0001-9309-0102]{Sergio B.~Fajardo-Acosta}
\affiliation{NASA Exoplanet Science Institute, Caltech/IPAC, 1200 E. California Blvd., Pasadena, CA 91125, USA}

\author[0000-0002-3853-7327]{Rachel B. Fernandes}
\altaffiliation{President's Postdoctoral Fellow}
\affil{Department of Astronomy \& Astrophysics, 525 Davey Laboratory, The Pennsylvania State University, University Park, PA 16802, USA}
\affil{Center for Exoplanets and Habitable Worlds, 525 Davey Laboratory, The Pennsylvania State University, University Park, PA 16802, USA}

\author[0000-0001-9800-6248]{Elise Furlan}
\affiliation{NASA Exoplanet Science Institute, Caltech/IPAC, 1200 E. California Blvd., Pasadena, CA 91125, USA}

\author[0000-0002-0388-8004]{Emily A. Gilbert}
\affiliation{Jet Propulsion Laboratory, California Institute of Technology, 4800 Oak Grove Drive, Pasadena, CA 91109, USA}

\author{Erica Gonzales}
\affiliation{Department of Astronomy and Astrophysics, University of California, Santa Cruz, CA 95064, USA}

\author[0000-0002-9903-9911]{Kathryn V.~Lester}
\affil{Mount Holyoke College, South Hadley MA 01075, USA}

\author[0000-0003-2527-1598]{Michael B.~Lund}
\affiliation{NASA Exoplanet Science Institute, Caltech/IPAC, 1200 E. California Blvd., Pasadena, CA 91125, USA}

\author[0000-0003-0593-1560]{Elisabeth C. Matthews}
\affiliation{Max-Planck-Institut f\"ur Astronomie, K\"onigstuhl 17, 69117 Heidelberg, Germany}

\author[0000-0001-7047-8681]{Alex S. Polanski}
\affil{Department of Physics and Astronomy, University of Kansas, Lawrence, KS 66045, USA}

\author[0000-0001-5347-7062]{Joshua E.~Schlieder}
\affil{NASA Goddard Space Flight Center, Greenbelt, MD 20771, USA}

\author[0000-0002-0619-7639]{Carl Ziegler}
\affiliation{Department of Physics, Engineering and Astronomy, Stephen F. Austin State University, 1936 North St, Nacogdoches, TX 75962, USA}

\begin{abstract}

M~dwarfs have become increasingly important in the detection of exoplanets and the study of Earth-sized planets and their habitability. However, 20$-$30\% of M~dwarfs have companions that can impact the formation and evolution of planetary systems. We use high-resolution imaging and Gaia astrometry to detect stellar companions around M~dwarf exoplanet hosts discovered by TESS and determine the projected separation and estimated stellar masses for each system. We find 47 companions around 216 M~dwarfs and a multiplicity rate of $19.4\pm2.7$\% that is consistent with field M~dwarfs. The binary projected separation distribution is shifted to larger separations, confirming the lack of close binaries hosting transiting exoplanets seen in previous studies. 
We correct the radii of planets with nearby companions and examine the properties of planets in M~dwarf multi-star systems. We also note three multi-planet systems that occur in close binaries ($\lesssim 50$~au) where planet formation is expected to be suppressed.

\end{abstract}


\section{Introduction} \label{sec:intro}

\nocite{2024AJ....167...56C} 

M~dwarfs are the most abundant stars in the Galaxy, dominating the nearby stellar population and accounting for $60-75$\% of all stars within 10 parsecs \citep{2006AJ....132.2360H, 2021A&A...650A.201R}. The prevalence of M~dwarfs, as well as their low masses, small sizes, 
and cool temperatures, have made them prime targets for exoplanet searches. In transiting exoplanet surveys, the small radii of M~dwarfs result in large planet-to-star radius ratios that facilitate the detection of transits and the discovery of small planets. Similarly, planets around M~dwarfs are attractive targets for atmospheric characterization and habitability based on their signal-to-noise compared to other systems and the proximity of the habitable zone to the star \citep{2020ApJ...891...58S}. The Transiting Exoplanet Survey Satellite (TESS) mission \citep{2015JATIS...1a4003R}, launched in 2018, is focused on detecting exoplanets around nearby bright stars and includes tens of thousands of M~dwarfs in its target list. Thus far more than 120 confirmed exoplanets have been detected around M~dwarfs using TESS\footnote{\url{https://exoplanetarchive.ipac.caltech.edu/}, as of 23 May 2024}, enabling improved constraints on exoplanet occurrence rates in low mass stars as well as the detection of giant planets \citep[e.g.,][]{2023MNRAS.521.3663B} and planets around mid-to-late M~dwarfs \citep[e.g.,][]{2023AJ....165..265M}.

In order to fully understand planet occurrence rates and statistical trends, however, the role of stellar multiplicity must be considered as stars commonly form in binaries. In transit surveys unidentified stellar companions contribute additional light that reduces the measured transit depth causing planet radii to be underestimated \citep{2015ApJ...805...16C} and Earth-sized planets to go undetected \citep[e.g.,][]{2021AJ....162...75L}, impacting both occurrence rates and survey completeness estimates \citep{2018AJ....155..244B, 2020ApJ...891...58S}. A binary companion is also expected to alter the formation and evolution of planetary systems. Stellar companions can decrease the masses and lifetimes of protoplanetary disks, impacting the likelihood of certain avenues of planet formation \citep{2015ApJ...799..147J, 2020MNRAS.496.5089Z}. Planets that form can also be perturbed by binary companions causing the migration and possible ejection of planets \citep{2013Natur.493..381K}. Nevertheless, simulations show dynamically stable planets can exist at semi-major axes within a few tenths of the binary separation \citep{1999AJ....117..621H}. Approximately 200 confirmed planets have been found in binary systems to date\footnote{\url{http://exoplanet.eu/planets_binary/}, \url{https://www.univie.ac.at/adg/schwarz/multiple.html}}, with many more unconfirmed candidates, emphasizing the need to understand the role stellar companions play in exoplanet formation and evolution. 

Observational efforts have largely concentrated on detecting companions to exoplanet host stars and calculating the binary fraction compared to field binaries. Such studies find that the wide binary population is consistent with the general stellar population \citep[e.g.][]{2014ApJ...795...60H, 2018AJ....156...83Z, 2018AJ....156...31M, 2021FrASS...8...16F, 2023AN....34430055M, 2024MNRAS.527.3183M}, but that close binaries infrequently host exoplanets \citep{2016AJ....152....8K, 2020AJ....159...19Z,2021AJ....162..192Z,2021MNRAS.507.3593M}, resulting in binary separation distributions peaking at wider distances than field stars \citep{2021AJ....161..164H, 2021AJ....162...75L, 2022AJ....163..232C, 2024AJ....167..174C}. A dedicated search for planetary and stellar companions in nearby solar-type stars similarly found giant planet occurrence rates to be equivalent in single stars and wide binaries ($>100$ au), with only 0.04 planets per star in close binaries \citep{2021AJ....161..134H}.

Although the multiplicity rate of stars decreases as a function of decreasing mass \citep{2013ARA&A..51..269D}, with multiplicity rates for field M~dwarfs determined to be $20 - 30$\% \citep[e.g.][]{2015MNRAS.449.2618W, 2019AJ....157..216W,2024AJ....167..174C}, the abundance of M~dwarfs and their importance for discovering and characterizing Earth-sized planets makes it necessary to understand the impact stellar companions have on their planetary systems. Planets discovered by TESS give us the opportunity to assess the multiplicity of M~dwarf exoplanet hosts, as their proximity allows follow-up observations sensitive to companions within a few au of the host star, at a level not possible with Kepler or K2 because of the larger distances of the M~dwarfs that do exist in those samples. For the M~dwarfs in their TESS Objects of Interest (TOI) survey, \citet{2020AJ....159...19Z, 2021AJ....162..192Z} found tentative evidence of fewer close binaries in planet hosting systems. More directly, \citet{2022AJ....163..232C} explored the multiplicity and orbital distribution of 58 M~dwarf TOIs, suggesting that planet-hosting M~dwarfs host fewer close-in stellar companions, similar to what has been shown for solar-type stars. This was confirmed for planet hosting M~dwarfs within 15pc in \citet{2024AJ....167..174C}.

In this paper we present observations of more than 200 M~dwarf TOIs to detect stellar companions and investigate the binary parameters for M~dwarfs hosting short period transiting planets. Our high-resolution imaging observations are described in Section~\ref{sec:obs} and the companion detections, including common proper motion analysis using Gaia, are detailed in Section~\ref{sec:companions}. The results allow us to explore the multiplicity of M~dwarf planet hosts in terms of the mass ratios of the binary star systems and the companion separation distribution in Section~\ref{sec:results}.  We then discuss the multiplicity fraction of M~dwarf transiting exoplanet host stars and examine the distribution of planet properties in multi-star systems in Section~\ref{sec:discussion}.

\section{Sample \& Observations} \label{sec:obs}

\nocite{https://doi.org/10.26134/exofop5}

\begin{figure}[t]
    \centering
    \includegraphics[width=0.5\textwidth]{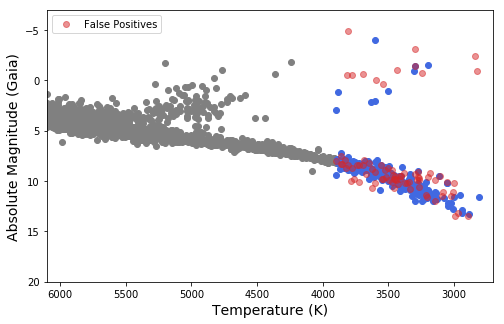}
    \caption{H-R Diagram of TESS TOI candidates. Grey symbols are randomly chosen F, G, K stars taken from the TOI list and plotted here to give a perspective for their locations in an H-R Diagram. Red dots are M star TOIs which have already been deemed to be false positives, while blue dots show the location of the remaining M star TOI candidates. The blue dots above the main sequence are deemed to be giants and are removed from any subsequent analysis.}
    \label{fig:hrdiagram}
\end{figure}

TOIs with a stellar effective temperature of 3900\,K or less in the TESS Input Catalog (TIC) were selected from the TESS Exoplanet Follow-up Observing Program (ExoFOP) website\footnote{\url{https://exofop.ipac.caltech.edu/tess/}; as of 27 January 2022.}, which resulted in a list of 308 potential M~dwarf exoplanet host stars. The absolute magnitude, determined from the apparent G magnitude and inverse parallax from Gaia DR3 \citep{2021A&A...649A...1G}, and  effective temperature of the host stars are plotted in blue in Figure~\ref{fig:hrdiagram}. As TOIs are preliminary planet candidates that require follow-up and validation efforts to be confirmed, some portion of the TOIs are later determined to be false positives or false alarms. Planetary transit false positives are often caused by spatially nearby eclipsing binaries or incorrect stellar radii (too small) thereby increasing the ``planet"  radius into the stellar regime. Sixty-five of the TOIs have been classified as false positives\footnote{as of 8 February 2024}, shown in red in Figure \ref{fig:hrdiagram}, and have been excluded from our sample. We also remove thirteen ambiguous planet candidates (APC) as many of them have large planet radii that are more likely to be eclipsing binaries than true planets \citep{2021MNRAS.507.3593M}. Such systems can lead to an excess of stellar companions when examining the multiplicity of host stars as $\sim$96\% of very close binaries (P~$<$~3~d) have a wide tertiary companion \citep{2006A&A...450..681T, 2020ApJ...902..107L}. Eight of the APCs have already been identified as binary stars via spectroscopy or the Gaia DR3 catalog of non-single stars \citep{2023A&A...674A...1G}. 
In Figure \ref{fig:hrdiagram} we also see that nine of the stars in our sample remain above the main sequence in the H-R diagram location 
associated with Luminosity III giants. We remove these likely giants to limit our assessment of planetary and stellar companions to main sequence M~dwarfs. Our final sample therefore consists of a total of 221 M~dwarfs with confirmed or candidate transiting exoplanets. Table~\ref{tab:example} lists the TOI, TIC, and Gaia designations for all M~stars, including those excluded from the final sample, and indicates the planet status and whether a stellar companion was detected. Any companions detected with high-resolution imaging around TOIs that were subsequently removed from our sample are reported in the Appendix for completeness.

To assess the stellar multiplicity of these systems, we collected high-resolution imaging observations of as many of the TOIs as possible. Such observations are routinely conducted to detect nearby objects that contaminate TESS lightcurves but are not resolved in the TIC or by ground based photometry. In the following sections we describe the observations conducted by our team and collaborators using speckle interferometry and adaptive optics, complementary techniques that can detect nearby stellar companions in the optical and infrared, respectively, and report imaging data from other groups available on ExoFOP.
 
\begin{deluxetable*}{lcccc}
\tabletypesize{\footnotesize}
\tablewidth{0pt}
\tablecaption{M~star TOIs with TIC and Gaia DR3 designations}\label{tab:example}
\tablehead{\colhead{TOI} & \colhead{TIC} & \colhead{Gaia ID} & \colhead{Planet Status} & \colhead{Detected Companions}} 
\startdata 
  203 & 259962054 & DR3 4647534190597951232 & FP  &       \\
  206 &  55650590 & DR3 4665277593852295424 &     &       \\
  210 & 141608198 & DR3 4650160717726370816 &     &       \\
  212 & 206609630 & DR3 4717055692347031296 & APC & HR    \\
  218 &  32090583 & DR3 4667466549703138176 &     & G     \\
\enddata
\tablecomments{A planet status of FP, APC, or g (likely giant) indicates the TOI was excluded from our sample. Stellar companions are identified by their detection method of high-resolution imaging (HR) and/or Gaia common proper motions (G). Companions found with high-resolution imaging but not bound to the TOI are marked with `U'. The table includes companions detected with high-resolution imaging around TOIs removed from our sample, however, not all such TOIs have been searched for companions. Table \ref{tab:example} is published in its entirety in the electronic edition of the {\it Astrophysical Journal}.  A portion is shown here for guidance regarding its form and content.}
\end{deluxetable*}

\subsection{Speckle Interferometry}
We observed 148 of the M~dwarf TOIs with speckle interferometry between 2019 January $-$ 2024 January. Most of the observations were obtained with the `Alopeke and Zorro speckle imagers \citep{2021FrASS...8..138S} at the Gemini 8.1~m North and South telescopes, respectively, while some additional TOIs were observed using the NESSI speckle imager at the 3.5~m WIYN telescope \citep{2018PASP..130e4502S}. A majority of the observations were part of the general TESS follow-up conducted by the NASA High-Resolution Speckle Interferometric Imaging Program \citep{2021FrASS...8...10H}, which observes planet candidates using the latest version of the TOI catalog. In addition, M star TOIs without any high-resolution imaging were specifically targeted during the 2021B - 2023A observing seasons. 

Each speckle observation consists of one or more sets, depending on the magnitude of the target star, of 1000 $\times$ 60~ms exposures taken in filters centered at 562 ($\Delta \lambda = 54$)~nm and 832 ($\Delta \lambda = 40$)~nm simultaneously. A bright single star is observed immediately before or after each target for point source calibration. `Alopeke and Zorro have plate scales of 0\farcs0096 per pixel and fields of view of $6.7''$, though companion detection is limited to a field of view of $\sim 2.4''$ as speckles become decorrelated beyond this limit. NESSI similarly detects companions within $\sim 2.4''$ and has a plate scale of 0\farcs0182 per pixel.

The data are reduced using a custom pipeline \citep{2011AJ....141...45H, 2011AJ....142...19H} that calculates the power spectrum for each target by taking the Fourier transform of the summed autocorrelation of each set of images and dividing it by the power spectrum of the point source. If a companion is detected, the pipeline calculates the separation, magnitude difference, and position angle of the companion relative to the target star. The pipeline is also used to produce reconstructed images of the target star via bispectrum analysis and $5\sigma$ sensitivity curves that correspond to the faintest companions detectable at a given separation (see Figure~\ref{fig:spkcomps}). The reconstructed images and detection limit curves for all TOI observations are available on the ExoFOP website. 

\subsection{Adaptive Optics}

We also observed 85 M~dwarf TOIs with near-infrared adaptive optics (AO) imaging using NIRC2 on the Keck-II 10~m telescope and PHARO on the 200 inch Hale telescope at Palomar Observatory from 2018 November $-$ 2022 August. These observations were conducted as part of the TESS high-resolution imaging follow-up program \citep[e.g.][]{2021FrASS...8...63S}, primarily focusing on small planets suitable for precision radial velocity follow-up as well as targets of interest to the broader community, though some M stars were specifically targeted at Palomar in 2022 for this study.

Observations at Keck used the NIRC2 instrument behind the Natural Guide Star AO system \citep{2000PASP..112..315W} and followed a standard three-point dither pattern, which avoids the noisier lower-left quadrant of the detector. The camera was used in the narrow-angle mode, resulting in a plate scale of 0\farcs09942 per pixel and a full field of view of $10''$.  Both narrow (e.g.~Br-$\gamma$, J$_{cont}$) and broad-band filters (e.g.~$K$) with central wavelengths near $2.2~\mu m$ were used depending on the NIR magnitude of the target and the observing conditions. 

Palomar observations made use of the PHARO instrument \citep{2001PASP..113..105H} behind the natural guide star AO system \citep{2013ApJ...776..130D}. Images were obtained in sets of 15 using a five-point dither pattern. The camera was in the narrow-angle mode with a plate scale of approximately 0\farcs025 per pixel and a full field of view of $\sim25''$. Observations were made in the Br-$\gamma$ ($\lambda_o = 2.1686; \Delta \lambda = 0.0326 \micron$) and/or H$_{cont}$ ($\lambda_o = 1.668; \Delta \lambda = 0.018\micron$) filter depending on the magnitude of the target, observing conditions, and whether a companion was detected. 

Science frames were sky-subtracted using a median average of the dithered science frames and flat-fielded using a median average of dark-subtracted flats taken each night. The reduced frames were then combined into a single image using an intrapixel interpolation that co-aligns and median-coadds the dithered frames while conserving flux. NIRC2 observations have a typical resolution of 0\farcs05 determined from the FWHM of the point-spread function (PSF), while Palomar observations have a typical resolution of 0\farcs1.

Contrast limits for each image were determined by injecting simulated sources in azimuthal increments at separations that are integer multiples of the central source's FWHM \citep[following][]{2017AJ....153...71F} and increasing the flux of each source until it was detected at the $5\sigma$ level with aperture photometry. The final contrast sensitivity as a function of separation was calculated by averaging all of the limits at that separation (see Figure~\ref{fig:AOcomps}). The reduced, combined images and contrast curves for all TOI observations are available on the ExoFOP website.

\subsection{ExoFOP Observations}
To augment our observations we inspected the additional high-resolution imaging data available on ExoFOP. These observations come from facilities around the world that produce better-than-seeing-limited images using techniques such as adaptive optics, speckle imaging, and lucky imaging in support of the TESS Follow-up Observing Program. The ExoFOP website, developed and maintained by the NASA Exoplanet Science Institute (NExScI), serves as a repository for such data products where users can upload their observations and results in order to pool resources and facilitate collaborations when investigating exoplanet candidates. Including these data increases the number of M star TOIs in our sample that have been searched for stellar companions by 41 and expands the parameter space searched for each TOI due to the different angular resolutions and photometric sensitivities of each technique. The TOIs in our sample have an average of $\sim$5 high-resolution observations per target, although 23 have no high-resolution imaging data available.

A majority of the observations retrieved from ExoFOP, including ten companion detections, were made with the HRCam speckle imager on the SOAR 4.1~m telescope \citep[see][]{2020AJ....159...19Z, 2021AJ....162..192Z}. Observations consist of two sets of 400 frames taken in the I band ($\lambda_o = 824; \Delta \lambda = 170$~nm) with a 6\farcs3 field of view and a plate scale of 0\farcs01575. For more details of the observation and data reduction procedure see \citet{2018PASP..130c5002T}. 

Additional speckle observations, including the detection of one companion, were made by the Speckle Polarimeter on the 2.5~m at the Caucasian Observatory of the Sternberg Astronomical Institute (SAI) of Lomonosov Moscow State University. For each target 4000 frames with 30 ms exposure times were obtained in the $I_c$ filter ($\lambda_o = 806$~nm) and used to determine the power spectrum. The full data reduction and calibration process is described in \cite{2017AstL...43..344S}. The Speckle Polarimeter has a plate scale of 0\farcs0206 per pixel and an angular resolution of 0\farcs08. 

The remaining observations retrieved from ExoFOP were used to help rule out the existence of stellar companions around the remaining TOIs as no additional companions were reported. For both the Gemini/NIRI \citep{2003PASP..115.1388H} and VLT/NaCo \citep{2003SPIE.4841..944L, 2003SPIE.4839..140R} adaptive optics observations, observing sequences consisted of nine images at nine dither positions in a grid pattern. For VLT/NaCo, the star was centered in the upper left quadrant for all dither positions to avoid detector problems in the other three quadrants, while for Gemini/NIRI it was centered in the center of the field of view. We used either the Br-$\gamma$ or Ks band filter for these observations, depending on the target magnitude, and customised the integration time for each observation based on the target magnitude. Both Gemini/NIRI images were processed similar to the Keck/NIRC2 data reduction described above: images were sky-subtracted using a median of the dithered science frames, flat-fielded, aligned based on the host star position in each frame, and co-added. Companions were identified through visual inspection and contrast sensitivities by injecting simulated sources into the images. Plate scales are 0\farcs0219 per pixel and 0\farcs0132 per pixel for Gemini/NIRI and VLT/NaCo, respectively.

High-resolution images were also obtained by the AstraLux instrument \citep{2008SPIE.7014E..48H} on the 2.2~m telescope of the Calar Alto observatory (Almer\'ia, Spain). Diffraction limited images were produced using lucky imaging, which obtains thousands of frames with exposure times below the atmospheric coherence time and selects only the $\sim$$10\%$ of frames with the highest Strehl ratios for processing. All images were obtained in the SDSSz bandpass ($\lambda_o = 909.7; \Delta \lambda = 137$~nm) with individual exposure times of $10 - 60$~ms and reduced with the AstraLux pipeline \citep{2008SPIE.7014E..48H}. More details of the AstraLux TESS observations, including how sensitivity limits are determined for each image, are described in \citet{2024A&A...686A.232L}. AstraLux has a resampled plate scale of 0\farcs02327 per pixel.

\section{Detected Companions} \label{sec:companions}

\subsection{High-resolution Imaging}\label{hrcomps}

\begin{figure}[]
    \centering
    \includegraphics[width=0.85\textwidth]{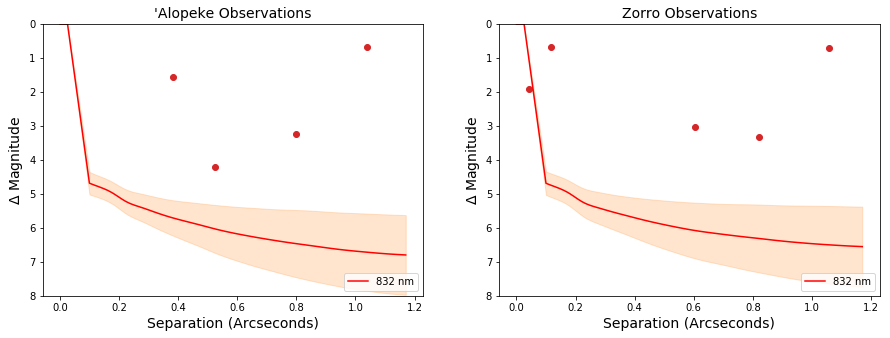}
    \caption{Magnitude difference as a function of angular separation for companions detected with speckle imaging in the 832~nm filter. Eight companions were detected around M~dwarf TOIs (one with both `Alopeke and Zorro). The shaded regions show the 5$\sigma$ detection limits for all observations in the 832 nm filter for `Alopeke and Zorro, respectively, with the average detection limit shown as a solid red line. We detect companions $<$ 1\farcs0 down to the diffraction limit of the telescope, confirming that our observation limits meet expectations.}\label{fig:specklecomps}
    \label{fig:spkcomps}
\end{figure}

Speckle observations detected eight companions to M~star TOIs with angular separations ranging from 0.04 - 1\farcs1. The magnitude difference as a function of angular separation for each companion measured in the 832~nm filter is shown in Figure~\ref{fig:specklecomps}, as not all companions were detected in the 562~nm filter. The one companion detected by NESSI was also observed with `Alopeke and is therefore not plotted separately. The $5\sigma$ detection limit curves for all M~star TOIs observed with `Alopeke and Zorro are also shown in Figure~\ref{fig:specklecomps}, with the average detection limit shown as a solid red line. We estimate our uncertainties in separation, position angle, and magnitude difference to be $0\farcs01$, 1.0\degr, and 0.5~mag, respectively, based on the uncertainties determined for speckle observations of M~dwarfs in \citet{2024AJ....167...56C}, the astrometric uncertainties for speckle imaging derived in \citet{2023AJ....166..166L}, and comparisons between companions in this work detected using multiple techniques.

\begin{figure}[b]
    \centering
    \includegraphics[width=0.85\textwidth]{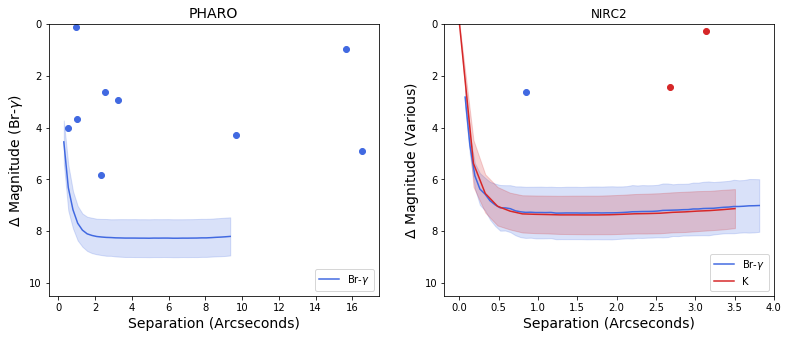}
    \caption{Magnitude difference as a function of angular separation in the Br-$\gamma$ filter for companions detected with AO imaging (blue points). K-band parameters (red points) are shown for two companions not observed in Br-$\gamma$. The 5$\sigma$ detection limits for all PHARO and NIRC2 observations are shown for the Br-$\gamma$ (blue) and K-band (red) filters, respectively, with solid lines highlighting the average detection limits. The AO observations complement our speckle observations as they have deeper contrast limits at wider separations, expanding our parameter space for high-resolution detections.}\label{AOcomps}
    \label{fig:AOcomps}
\end{figure}

Twelve companions with separations ranging from 0.5 - 16\farcs5 were detected using AO observations. Two of the companions were also detected via speckle imaging and have measured angular separations consistent within $0\farcs05$. The magnitude difference versus angular separation in Br-$\gamma$ (blue) or K-band (red) for each companion detected with AO is plotted in Figure~\ref{fig:AOcomps}, as well as the sensitivity curves for all PHARO and NIRC2 observations and the mean sensitivity curve for each instrument and filter (solid lines). Average uncertainties in separation, position angle, and magnitude difference of the 12 companions are $0\farcs008$, 0.8\degr, and 0.02~mag, respectively.

The high-resolution imaging data extracted from ExoFOP contains 13 companions around 11 TOIs. Eight of the companions were not observed or not detected in our observations, giving us a total of 26 companions around 24 TOIs. The SOAR speckle program observed eight TOIs with companions that were also observed by our speckle program. The three companions detected by both programs have separations and position angles consistent within $0.008$'' and $3$\degr, respectively. One companion was detected only with Zorro as it is below the diffraction limit of SOAR in I-band (0\farcs05) and four were detected only with HRCam, two of which are outside our speckle field of view. 

The remaining two companions detected only with HRCam are close companions that orbit the target TOI as part of a hierarchical triple system. TOI-2221 is the M type variable star AU Mic, which hosts two planets \citep{2020Natur.582..497P, 2021A&A...649A.177M}, and has a high statistical probability of being part of a very wide triple system with the BC components approximately $\sim$1.3\degr~away \citep{2011ApJS..192....2S}. According to the Washington Double Star Catalog \citep[WDS;][]{2001AJ....122.3466M} the B and C components were separated by 2\farcs1 and a position angle of 146\degr~in 2015, with the separation and position angle gradually decreasing throughout the twentieth century, consistent with the HRCam measure from 2020 (1\farcs76, 128\degr). Since TOI-2221 has two high probability stellar companions, as well as two confirmed planets, we keep the system in our sample and report the separations of B and C as 4680\farcs5 ($\Delta$m = 1.74) and 4682\farcs2 ($\Delta$m = 2.72), respectively. Similarly, TOI-455 is a known hierarchical triple consisting of three M~dwarfs with the BC subsystem approximately 7'' away from the primary as of 2017, while the B and C components have a measured separation of 1\farcs0 in 2019 and 1\farcs2 in 2020 in the WDS. Therefore, the companion observed with HRCam in 2019 is likely the C component measured relative to the B component. TOI-455 also has two confirmed planets \citep{2019AJ....158..152W, 2022AJ....163..168W}, so we keep the system in our sample but adopt $7''$ and $8''$ as the separations of the B and C stellar companions with respect to the primary and calculate delta magnitude values from the component magnitudes reported in \citet{2019AJ....158..152W}. As a result, we have a total of 28 companions around 24 TOIs in our sample.

\begin{figure}
    \centering
    \includegraphics[width=0.45\textwidth]{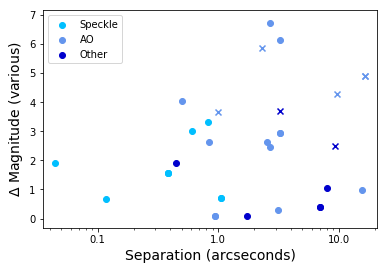}
    \caption{Magnitude difference as a function of angular separation for all companions detected with high-resolution imaging (excluding TOI-2221). Companions detected in more than one technique are only plotted once, with properties adopted (in order) from AO imaging, speckle, or the  observations available on ExoFOP (other). Note that the magnitude differences were measured in different filters (Br-$\gamma$, K, I, or 832 nm) and have not been placed on a uniform scale. Likely unbound companions are shown as $\times$'s (see Section~\ref{boundcomps}). The ExoFOP companions fit well within the established parameter space, increasing the number of TOIs with detections by high-resolution imaging.} \label{fig:HRcomps}
\end{figure}

All companions detected with high-resolution imaging are listed in Table~\ref{tab:hirescomps}, with the target TOI, instrument, filter, angular separation, position angle, magnitude difference, and date of observation. The magnitude difference as a function of angular separation for each companion is plotted in Figure~\ref{fig:HRcomps} (excluding TOI-2221). Companion parameters in Figure~\ref{fig:HRcomps} are color-coded by the detection technique, which are adopted from AO imaging, if available, then speckle imaging, and finally the additional observations from ExoFOP. Magnitude differences are measured in different filters for each technique and have not been transformed to a common wavelength or filter. For companions observed with multiple techniques, the mean standard deviation of the measured separations and position angles are 0.01$''$ and 1.5\degr, respectively.

\startlongtable
\movetabledown=10mm
\begin{deluxetable*}{lcccccc}
\tabletypesize{\footnotesize}
\tablewidth{0pt}
\tablecaption{Companions observed with high-resolution imaging}
\tablehead{
\colhead{TOI} & \colhead{Instrument} & \colhead{Filter} & \colhead{Separation ($''$)} & \colhead{Position Angle ($\degr$)} & \colhead{$\Delta$ Magnitude} & \colhead{Date}} 
\startdata 
256\tablenotemark{b} &                                          PHARO  &                       Br-$\gamma$ &                           16.52  &                        336  &                       4.8987  &                                        2018-12-22  \\
455\tablenotemark{a} &                                          HRCam  &                             I  &                          7  &                      314  &                          0.39  &                                        2019-08-12  \\
  455\tablenotemark{a} &                                          HRCam  &                             I  &                          8  &                      314  &                       1.05     &                                        2019-08-12  \\
  457 &                                'Alopeke  &                          832  &             1.039  &          222.9  &             0.67 &                2019-10-14  \\
   &                                Zorro  &                          832  &             1.056  &           44.8 &            1.06  &                2020-01-09  \\
   &                                Zorro  &                          832  &             1.059  &           44.7  &            0.71  &                2020-01-14  \\
654\tablenotemark{b} &                                          PHARO  &                Br-$\gamma$ &                     9.65 &                   120 &               4.2842  &                                        2019-06-13  \\
                       &                                          PHARO  &                Hcont  &                     9.178  &                   115  &               4.1679  &                                        2019-06-13  \\
  737 &                                NIRC2  &           Br-$\gamma$ &             0.842 &                   46  &          2.615 &                            2019-06-10  \\
  &                                NIRC2  &           Jcont  &             0.845  &                   46  &          2.6418  &                            2019-06-10  \\
  &                                'Alopeke  &            832  &             0.798  &                   47.7  &          3.25  &                             2020-02-16  \\
864 &                                          Zorro  &                  562  &            0.04 &                       51.3 &       0.35  &                                        2022-01-13  \\
      &                                          Zorro  &                  832  &            0.044  &                     45.8  &      1.92  &                                        2022-01-13  \\
 1215 &                                          HRCam  &                             I  &                          1.7263  &                       84.7  &                          0.1  &                                        2019-12-14  \\
1450 &                                          PHARO  &                Br-$\gamma$  &                    3.253  &               324.5  &                 2.936  &                                        2021-09-20  \\
      &                                          PHARO  &                Hcont  &                     3.248  &               324.7  &                 3.111  &                                        2021-09-20  \\
 1452 &                                          NIRC2  &                K  &                    3.134  &               0.36 &                 0.284  &                              2020-05-28  \\
 1634 &                                          PHARO  &                Br-$\gamma$  &                    2.538 &                      270.7  &                 2.619 &                                        2021-09-19  \\
      &                                          PHARO  &               Hcont  &                    2.535  &                      270.7  &                 2.713  &                                        2021-09-19 \\
1745\tablenotemark{b} &                                          PHARO  &                Br-$\gamma$   &                     0.98  &                     7.5  &   3.752  &                            2021-08-24  \\
                       &                                          PHARO  &                 Hcont  &                     0.98  &                     7.5  &   3.716  &                             2021-08-24  \\
                       &                                          PHARO  &                Br-$\gamma$  &                      1.003  &                      10  &   3.673  &                            2021-11-11  \\
                       &                                          PHARO  &                 Hcont  &                     1.003  &                   10  &  3.479  &                           2021-11-11  \\
 1746 &                                          PHARO  &                Br-$\gamma$  &                    0.943  &               173.8  &                 0.098  &                                        2021-11-11 \\
      &                                          PHARO  &                Hcont  &                    0.933  &                173.5  &               0.101  &                                        2021-11-11  \\
 1883 &                                          PHARO  &                       Br-$\gamma$  &                           15.66  &                      196.5  &                       0.9727  &                                        2021-02-24  \\
 1899\tablenotemark{b} &                                          PHARO  &                Br-$\gamma$  &                     2.31 &               355.4  &                  5.85&                                        2021-09-20  \\
                       &                                          PHARO  &                Hcont  &                     2.301  &               355.1  &                  5.885  &                                        2021-09-20  \\
 2221\tablenotemark{a} &                                          HRCam  &                             I  &                          4680.5 &                      212  &                          1.74  &                                        2020-12-03  \\
 2221\tablenotemark{a} &                                          HRCam  &                             I  &                          4682.2  &                      211.9  &                          2.72  &                                        2020-12-03  \\
 2267 &                            Speckle Polarimeter  &                             I  &                           0.408  &                     285.16  &                          1.8  &                                        2020-10-25  \\
  &                            'Alopeke  &                             832  &                           0.384  &                     279.7  &                          1.55  &                                       2021-12-09  \\
2384   &                                          Zorro  &                             832  &   0.822  &   349.3 &       3.32       &                  2022-11-10         \\
 2455 &                              HRCam  &             I  &            0.5146  &          119.2  &               3.0  &                            2021-03-01  \\
  &                                   PHARO  &              Br-$\gamma$  &            0.505  &           120 &               4.03  &                             2021-09-19 \\
  &                                   PHARO  &              Hcont  &            0.506  &           120.2  &               2.388  &                             2021-09-19 \\
  &                                   NESSI  &              832  &            0.504  &           118.87  &               2.92  &                             2021-10-29  \\
  &                                   'Alopeke  &              832  &            0.497  &           123.9  &               4.34  &                             2021-10-16  \\
  &                                   'Alopeke  &              832 &            0.525  &           121.3  &                4.22  &                             2022-02-12  \\
 2781 &                                          HRCam  &                             I  &                          0.4471  &                       94.8  &                          1.9  &                                        2021-10-03  \\
3494 &                                          HRCam  &                             I &                          0.5927  &                      293.1  &                          2.9  &                                        2021-07-15  \\
      &                                          HRCam  &                             I  &                         0.6017  &                      293.1  &                          2.9  &                                        2022-06-13  \\
      &                                          Zorro  &                             832  &                         0.605  &                     293.1  &                          3.02  &                                        2023-04-07  \\
4325\tablenotemark{b} &                                          HRCam  &                             I  &                   3.2583  &                233.5  &                      3.7  &                                        2021-10-18  \\
 4325\tablenotemark{b} &                                          HRCam  &                             I  &                  9.2699  &               214.6  &                     2.5 &                                        2021-10-18  \\
 4349 &                                 NIRC2    &                          K  &                  2.681   &               347.1   &                    2.45  &                            2021-08-28  \\
 &                                  HRCam   &                           I  &                  2.7053    &              12.4   &                     2.9  &                             2021-10-01  \\
 4446\tablenotemark{c} &                                   NIRC2  &                          K  &                  2.725   &                7.6   &                   6.72  &                            2012-06-24  \\
 4446\tablenotemark{c} &                                   NIRC2  &                          K  &                  3.298    &                68.4    &                   6.151   &                            2012-06-24  \\
 4889 &                                   Zorro  &                          562  &                  0.119   &                226.2    &                   3.12   &                            2022-03-21  \\
      &                                   Zorro  &                          832  &                  0.118    &               222.9    &                   0.67   &                            2022-03-21  \\
 &                                   HRCam  &                          I  &                 0.1268    &                47.5    &                   0.0   &                            2022-06-11  \\
\enddata
\tablenotetext{a}{Companion parameters on ExoFOP refer to the outer components of a triple system. Here we show the companion parameters with respect to the primary star. See Section~\ref{hrcomps} for details.}\tablenotetext{b}{Unbound, see Section \ref{boundcomps}}
\tablenotetext{c}{KOI-245, observed by Kraus et al.~2016, ExoFOP observations extracted from Furlan et al.~2017.}
\end{deluxetable*}
 \label{tab:hirescomps}

\vspace{-2cm}

\subsection{Common Proper Motions}\label{gaiacomps}

We also searched for wide companions to the M~dwarf TOIs using Gaia DR3 \citep{2023A&A...674A...1G} and code adapted from \cite{2021MNRAS.506.2269E} to identify common proper motion (CPM) pairs. The original code performs a volume-limited search for all potential CPM binaries in the Gaia archive. It applies three criteria to determining potential binaries: the projected separation is under one parsec, the parallaxes must be similar within 3$\sigma$, and the proper motions align with Keplerian orbits. After these criteria are applied, the code screens for possible higher-order systems and globular clusters. 

For this work, we cross-matched our sample of M~dwarfs with Gaia DR3 and performed companion searches in 10-arcminute cones around each TOI with a G-magnitude and nonzero parallax. All three requirements for identifying binaries remained the same as \citet{2021MNRAS.506.2269E}, which resulted in 53 TOIs with common proper motion companions. We did not screen for higher-order systems or globular clusters, resulting in large numbers of potential common proper motion companions for several of the TOIs. To eliminate unrealistic candidates we used the estimated stellar masses of the M~dwarf and companion (see \S \ref{sec:starprop}) to determine the maximum separation as a function of binding energy, as in \citet{2021A&A...650A.190G}, and removed companions beyond that separation. We also eliminated potential companions with low fractional parallax uncertainties, cutting companions with a combined {\tt\string parallax\_over\_error}~$< 5$. 

After applying these cuts TOI-5090 still returned 14 potential common proper motion companions. While listed as a known planet host in ExoFOP, it is a low mass eclipsing binary consisting of an M3.9 star and a 54 M$_\mathrm{J}$ brown dwarf and a high probability member of the Praesepe open cluster \citep{2017ApJ...849...11G}, so we exclude it from our sample. In addition, TOI-2496 appears in the Gaia DR3 catalog of non-single stars indicating it is a binary rather than the host of a 23~R$_\earth$ planet, and is also excluded from our analysis.

\begin{figure}[h!]
    \centering
    \includegraphics[width=0.95\textwidth]{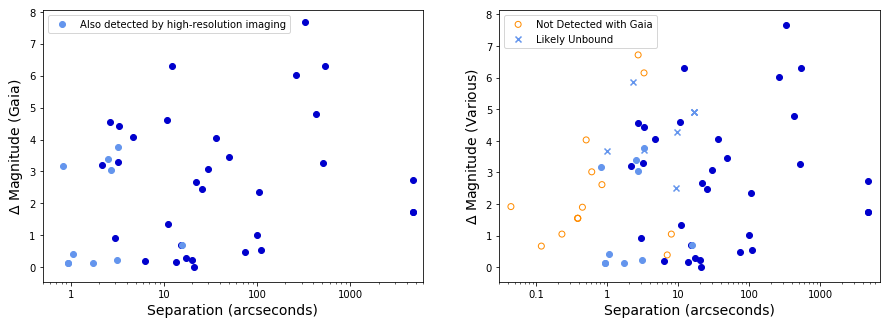}
    \caption{\textit{Left}: Magnitude difference as a function of angular separation for companions detected via common proper motion and parallax using Gaia DR3. Companions also detected with high-resolution imaging are shown in light blue. \textit{Right}: Companions detected via Gaia DR3 and the CPM companions of TOI-2221 are plotted in blue, with companions detected using high-resolution imaging and not also detected using Gaia shown in orange. Open circles are companions unresolved by Gaia, or with no proper motion and/or parallax data, which are assumed to be bound (see text for details). Blue $\times$'s show companions present in Gaia with parallax and proper motion data indicating they are not physically bound. Each companion is only plotted once, with overlapping companions a result of the magnitude differences measured in different filters. Comparing the separation ranges of the high resolution and CPM detections demonstrates that while there is overlap, they provide complementary samples.}\label{fig:Gaiacomps}
\end{figure}

The calculated separation, position angle, and magnitude difference for the remaining 38 CPM companions around 35 TOIs are shown in Table~\ref{tab:gaiacomps}, with the Gaia magnitude, parallax, and proper motions given for each TOI (first row) and companion (second row). Average uncertainties on the derived astrometry are approximately 1 mas in angular separation, 0.05\degr~in position angle, and 0.005 for the magnitude difference. The separation and magnitude difference of each companion is plotted in Figure~\ref{fig:Gaiacomps}. The detected companions have separations that range from $0.8 - 540''$, corresponding to projected separations of $40 - 97,000$ au. Nine of the companions were also identified with high-resolution imaging and are shown in light blue in Figure~\ref{fig:Gaiacomps}. The measured separations of these companions range from $0.8 - 16''$ and differ by 0\farcs04 on average.

Many of the CPM companions identified here were also reported in the survey of stellar companions to TOIs using Gaia data by \citet{2020AN....341..996M, 2021AN....342..840M} and \citet{2022AN....34324017M, 2023AN....34430055M} as well as the catalog of spatially resolved binaries from Gaia by \cite{2021MNRAS.506.2269E}. A majority of the companions not identified in these surveys have angular separations of a few hundred arcsec and projected separations $\gtrsim 25,000$~au. In addition, \cite{2021MNRAS.506.2269E} only identified binary companions, omitting any stars with more than one common proper motion companion. References to the relevant papers are shown in the last column of Table~\ref{tab:gaiacomps}.

\startlongtable
\movetabledown=10mm
\begin{deluxetable*}{lhcccccccc}
\tabletypesize{\scriptsize}
\tablewidth{0pt} 
\tablecaption{Companions detected using proper motions and parallaxes \label{tab:gaiacomps}}
\tablehead{
\colhead{TOI} & \nocolhead{TIC} & \colhead{Sep. ($''$)} & \colhead{Pos. Angle ($\degr$)} & \colhead{$\Delta$ Mag. (Gaia)} & \colhead{Mag. (Gaia)} & \colhead{Parallax (mas)} & \colhead{PM RA (mas)} & \colhead{PM Dec (mas)} & \colhead{Add.~Ref}}
\startdata 
218 &   32090583 &    13.53 & 226.09 &              0.15 &   14.63 &    19.03$\pm$0.02 &   139.03$\pm$0.02 &   191.18$\pm$0.02 &  M20,E21 \\
 &   32090583 &     &         &        &   14.79 &    19.03$\pm$0.02 &  139.90$\pm$0.02  &  191.23$\pm$0.02  &  \\
277\tablenotemark{a} &  439456714 &    17.25 & 352.37 &               0.30 &   12.78 &    15.41$\pm$0.02 &  -115.33$\pm$0.02 &  -249.42$\pm$0.01 &  M20,E21 \\
 &  439456714 &     &          &       &   12.47 &    15.44$\pm$0.02 &  -113.72$\pm$0.01 & -248.93$\pm$0.02  &   \\
457\tablenotemark{b} &   89256802 &     1.03 & 222.01 &               0.42 &   15.63 &     7.98$\pm$0.07 &    83.50$\pm$0.03 &    47.84$\pm$0.05 &  M20,E21 \\
 &   89256802 &      &       &          &   16.05 &     7.95$\pm$0.09 &   78.78$\pm$0.07  &   51.77$\pm$0.05  &   \\
468\tablenotemark{a} &   33521996 &   329.51 & 256.01 &               7.67 &   14.34 &     5.89$\pm$0.02 &    -2.57$\pm$0.01 &     7.37$\pm$0.05 &          \\
 &   33521996 &    &         &        &   5.91 &  6.68$\pm$0.02    &   -2.49$\pm$0.01  &   7.47$\pm$0.01   &          \\
488 &  452866790 &    49.26 & 222.79 &               3.47 &   12.45 &    36.61$\pm$0.02 &  -403.20$\pm$0.02 &  -380.93$\pm$0.01 &  M20,E21 \\
 &  452866790 &     &       &          &   15.92 &    36.45$\pm$0.07 & -399.42$\pm$0.04  &  -380.74$\pm$0.05 &   \\
507 &  348538431 &    73.46 & 340.15 &               0.47 &   14.48 &     9.10$\pm$0.02 &    47.71$\pm$0.01 &   -15.27$\pm$0.01 &  M20,E21 \\
 &  348538431 &     &        &         &   14.95 &     9.12$\pm$0.02 &  47.69$\pm$0.01   &  -15.04$\pm$0.02  &   \\
756 &   73649615 &    11.09 & 163.20 &               1.34 &   13.68 &    11.61$\pm$0.02 &  -216.50$\pm$0.01 &    29.20$\pm$0.01 &  M20,E21 \\
 &   73649615 &     &         &        &   15.02 &    11.61$\pm$0.03 & -215.7$\pm$0.02  &  29.5$\pm$0.02   &   \\
762 &  178709444 &     3.20 & 239.54 &               3.29 &   14.93 &    10.12$\pm$0.02 &  -159.17$\pm$0.02 &   -24.78$\pm$0.02 &  M20,E21 \\
 &  178709444 &      &         &        &   18.22 &    9.79$\pm$0.14 &  -157.37$\pm$0.1 &  -24.38$\pm$0.1  &   \\
1215\tablenotemark{b} &  453260209 &     1.76 & 84.55 &               0.12 &   11.42 &    28.89$\pm$0.02 &  -131.35$\pm$0.02 &  -117.48$\pm$0.02 &  M20,E21 \\
 &  453260209 &      &        &         &   11.42 &    28.89$\pm$0.02 & -143.02$\pm$0.02  & -120.86$\pm$0.02  &   \\
1227\tablenotemark{a} &  360156606 &   511.83 & 270.53 &               3.26 &   15.22 &     9.90$\pm$0.02 &   -40.29$\pm$0.02 &   -10.81$\pm$0.02 &          \\
 &  360156606 &    &         &        &  11.96  &     8.54$\pm$0.65 & -42.09$\pm$0.54   & -8.72$\pm$0.55   &          \\
1450\tablenotemark{b} &  377293776 &     3.39 & 323.67 &              3.78 &   11.24 &    44.56$\pm$0.02 &    69.99$\pm$0.02 &    11.24$\pm$0.01 &  M20,E21 \\
 &  377293776 &      &       &          &   15.02 &    44.52$\pm$0.03 &  83.31$\pm$0.03   & 143.77$\pm$0.03    &   \\
1452\tablenotemark{b} &  420112589 &     3.18 & 0.03 &               0.23 &   13.60 &    32.78$\pm$0.01 &     7.80$\pm$0.01 &   -74.08$\pm$0.01 &  M20,E21 \\
 &  420112589 &      &       &          &   13.83 &    32.79$\pm$0.01 &   6.85$\pm$0.01   &  -82.22$\pm$0.01  &   \\
1634\tablenotemark{b} &  201186294 &     2.54 & 269.36 &               3.40 &   12.19 &    28.51$\pm$0.02 &    81.35$\pm$0.01 &    13.55$\pm$0.01 &  M20,E21 \\
 &  201186294 &      &        &         &   15.59 &    28.62$\pm$0.11 &  80.64$\pm$0.07   &   14.54$\pm$0.09  &   \\
1696 &  470381900 &   426.19 & 208.66 &               4.80 &   15.31 &    15.48$\pm$0.03 &    12.87$\pm$0.03 &   -19.05$\pm$0.02 &          \\
 &  470381900 &    &           &      &   15.31 &    9.89$\pm$1.94 &  13.99$\pm$0.89   &  -26.81$\pm$1.49  &          \\
1746\tablenotemark{b} &  232650365 &     0.93 & 171.41 &               0.15 &   13.80 &    22.69$\pm$0.11 &     7.40$\pm$0.13 &    55.38$\pm$0.11 &  M20,E21 \\
 &  232650365 &      &        &         &   13.80 &    22.64$\pm$0.04 &   9.37$\pm$0.04   &   53.82$\pm$0.06  &   \\
1763 &  441739871 &    15.11 & 66.61 &               0.71 &   13.94 &    11.35$\pm$0.01 &    -5.01$\pm$0.01 &   -38.44$\pm$0.01 &  M20,E21 \\
 &  441739871 &     &         &        &   14.65 &    11.34$\pm$0.02 &  -4.41$\pm$0.02   & -38.18$\pm$0.02   &   \\
1801 &  119584412 &   105.05 & 21.97 &              2.36 &   10.83 &    32.37$\pm$0.02 &  -204.89$\pm$0.02 &    41.84$\pm$0.02 &  M22,E21 \\
 &  119584412 &    &          &       &   13.19 &    32.37$\pm$0.02 &  -204.91$\pm$0.02 &   42.49$\pm$0.01  &   \\
1883\tablenotemark{b} &  348755728 &    15.75 & 194.94 &              0.69 &   14.50 &     8.51$\pm$0.02 &    30.78$\pm$0.02 &     2.84$\pm$0.02 &  M20,E21 \\
 &  348755728 &       &        &         &  15.19  &    8.58$\pm$0.04  &   30.58$\pm$0.03  &   3.15$\pm$0.03   &   \\
1883 &  348755728 &   261.14 & 236.20 &               6.02 &   14.50 &     8.51$\pm$0.02 &    30.78$\pm$0.02 &     2.84$\pm$0.02 &          \\
 &  348755728 &    &         &        &   20.52 &     13.12$\pm$2.56 &  18.84$\pm$1.70   &  7.04$\pm$1.40    &          \\2068 &  417931300 &    19.99 & 65.49 &               0.23 &   12.21 &    18.87$\pm$0.01 &  -197.94$\pm$0.01 &    -6.06$\pm$0.01 &  M21,E21 \\
 &  417931300 &     &        &         &   12.44 &    18.86$\pm$0.01 & -200.44$\pm$0.01  &  -7.29$\pm$0.01   &   \\
2072 &  900715901 &     2.99 & 130.52 &               0.92 &   12.70 &    25.63$\pm$0.02 &  -184.97$\pm$0.02 &   -87.24$\pm$0.02 &  M21,E21 \\
 &  900715901 &      &       &          &   13.61 &    25.55$\pm$0.02 & -184.59$\pm$0.02  & -92.03$\pm$0.01   &   \\
2084 &  441738827 &    12.25 & 191.19 &               6.29 &   14.39 &     8.75$\pm$0.02 &    47.73$\pm$0.02 &    36.76$\pm$0.01 &  M21,E21 \\
 &  441738827 &     &        &         &   14.65 &     8.79$\pm$0.75 &  49.1$\pm$0.56   &  38.89$\pm$0.67   &   \\
2094 &  356016119 &    10.71 & 215.71 &               4.60 &   13.43 &    19.91$\pm$0.01 &   -55.30$\pm$0.01 &    -0.13$\pm$0.01 &  M21,E21 \\
 &  356016119 &     &        &         &   18.03 &    20.06$\pm$0.12 & -54.38$\pm$0.11   &  1.27$\pm$0.12   &   \\
2205 &  260298958 &     2.15 & 83.39 &               3.19 &   16.13 &     2.41$\pm$0.03 &    -2.33$\pm$0.03 &    27.63$\pm$0.03 &      M21 \\ &  260298958 &      &       &          &   19.32 &     2.39$\pm$0.14 &  -2.12$\pm$0.15   &  27.47$\pm$0.13   &      \\
2205 &  260298958 &    21.83 & 255.41 &               2.66 &   16.13 &     2.41$\pm$0.03 &    -2.33$\pm$0.03 &    27.63$\pm$0.03 &      M21 \\
 &  260298958 &     &         &        &   18.79 &     2.39$\pm$0.14 &  -2.12$\pm$0.13   &   27.47$\pm$0.15  &       \\
2293 &   71347873 &     4.66 & 1.24 &               4.07 &   12.86 &    15.94$\pm$0.02 &    45.22$\pm$0.01 &  -120.70$\pm$0.01 &  M21,E21 \\
 &   71347873 &      &        &         &   16.93 &    16.08$\pm$0.06 &  45.11$\pm$0.05   &  -123.74$\pm$0.04 &   \\
2384\tablenotemark{b} &  382602147 &     0.84 & 350.29 &               3.16 &   14.39 &     5.32$\pm$0.04 &     9.26$\pm$0.04 &   -52.66$\pm$0.04 &  M21,E21 \\
 &  382602147 &      &        &         &   17.55 &     5.05$\pm$0.17 &   8.01$\pm$0.19   &  -51.04$\pm$0.44  &   \\
3397 &  154616309 &   536.57 & 268.28 &               6.30 &   14.29 &     5.50$\pm$0.02 &    14.50$\pm$0.01 &     9.73$\pm$0.01 &          \\
 &  154616309 &    &          &       &   20.59 &     12.09$\pm$2.24 &  3.48$\pm$1.77   &  10.89$\pm$1.1    &          \\3714 &  155867025 &     2.67 & 106.42 &               4.56 &   14.29 &     8.84$\pm$0.02 &    19.83$\pm$0.02 &   -70.76$\pm$0.01 &  M23,E21 \\
 &  155867025 &      &        &         &   18.85 &     8.85$\pm$0.23 &  18.8$\pm$0.11   &  -70.67$\pm$0.2   &   \\
3984 &   20182780 &     3.27 & 343.08 &               4.43 &   14.65 &     9.18$\pm$0.02 &   -48.95$\pm$0.01 &    42.65$\pm$0.02 &  M23,E21 \\
 &   20182780 &      &         &        &   19.08 &     9.06$\pm$0.2 &  -50.38$\pm$0.16  &  40.95$\pm$0.13   &   \\
4325 &  202185707 &    25.30 & 356.26 &               2.46 &   12.98 &    19.53$\pm$0.02 &    74.70$\pm$0.02 &  -169.87$\pm$0.02 &  M23,E21 \\
 &  202185707 &     &          &       &   15.44 &    19.51$\pm$0.04 &  74.78$\pm$0.03   & -170.67$\pm$0.03  &   \\
4336 &  166184428 &     6.30 & 302.79 &               0.19 &   12.25 &    44.53$\pm$0.04 &   151.81$\pm$0.03 &    68.40$\pm$0.02 &          \\
 &  166184428 &      &        &         &   12.43 &    44.55$\pm$0.04 &  150.41$\pm$0.03  &  71.66$\pm$0.03   &          \\
4336 &  166184428 &    98.50 & 197.59 &               1.01 &   12.25 &    44.53$\pm$0.04 &   151.81$\pm$0.03 &    68.40$\pm$0.02 &          \\
 &  166184428 &     &         &        &   13.26 &    44.50$\pm$0.02 &  151.99$\pm$0.01  &   71.80$\pm$0.01  &          \\
4349\tablenotemark{b} &  313889049 & 2.68 & 12.35 &                3.04 &   12.66 &    13.89$\pm$0.02 &   278.22$\pm$0.02 &   105.50$\pm$0.02 &  M23,E21 \\
 &  313889049 &      &       &          &   15.70 &    13.86$\pm$0.11 &  280.63$\pm$0.08  &  108.14$\pm$0.09  &   \\
4642\tablenotemark{a} &  336961891 &   110.77 & 238.42 &               0.54 &   13.27 &    42.44$\pm$0.02 &    92.00$\pm$0.02 &  -122.04$\pm$0.02 &      E21 \\
 &  336961891 &    &         &        &   12.73 &    42.53$\pm$0.03 &  94.02$\pm$0.02   & -120.39$\pm$0.03  &       \\
4668 &  142938659 &    20.91 & 194.95 &               0.02 &   15.16 &     9.43$\pm$0.02 &    35.72$\pm$0.02 &    -4.42$\pm$0.02 &          \\
 &  142938659 &     &        &         &   15.19 &     9.42$\pm$0.03 &  35.24$\pm$0.02   &  -4.31$\pm$0.02   &          \\
4858 &  262499797 &    36.34 & 35.89 &               4.06 &   16.05 &     5.04$\pm$0.03 &    54.19$\pm$0.03 &   -92.53$\pm$0.03 &      E21 \\
 &  262499797 &     &         &        &   20.11 &     4.41$\pm$0.51 &  51.91$\pm$0.52   & -93.23$\pm$0.49   &       \\
4991 &  247180220 &    30.01 & 150.17 &                3.08 &   15.75 &     2.91$\pm$0.04 &    10.80$\pm$0.03 &    -3.03$\pm$0.02 &      E21 \\
 &  247180220 &     &          &       &   18.83 &     2.29$\pm$0.23 &  10.09$\pm$0.14  &  -3.01$\pm$0.2   &       \\
\enddata
\tablenotetext{a}{Detected companion is brighter than the TOI. For these systems, the mass ratio is calculated using the brighter star as primary.}
\tablenotetext{b}{Companion also detected with high-resolution imaging.}
\tablecomments{References for previously identified CPM companions: M20 - \citet{2020AN....341..996M}, M21 - \citet{2021AN....342..840M}, M22 - \citet{2022AN....34324017M}, M23 - \citet{2023AN....34430055M}, E21 - \citet{2021MNRAS.506.2269E}}
\end{deluxetable*}

\vspace{-1cm} 
\subsection{Companion Assessment} \label{boundcomps}

TOIs with companions detected using high-resolution imaging were cross-matched with Gaia DR3 to evaluate whether the companions are bound or background stars. Six of the companions have proper motions and parallaxes that differ from the TOI host star by a factor of four or more, shown as $\times$'s in Figures~\ref{fig:HRcomps} and \ref{fig:Gaiacomps}, and are treated as single stars in our analysis. The remaining companions were not resolved by Gaia or had no parallax and/or proper motion measurements and were therefore assumed to be bound companions. This assumption is reasonable as these companions have separations less than $1''$, except the known triple TOI-455, and previous studies have found most sub-arcsecond companions are physically associated with the target star \citep[e.g.][]{2017AJ....153...25A, 2014ApJ...795...60H, 2018AJ....156...83Z}. While these studies examined stellar companions to Kepler Objects of Interest (KOIs), \citet{2017AJ....153...25A} determined only $14.5\%$ of companions with separations up to $4''$ are likely unbound and  \citet{2019AJ....157..211M} found the majority of companions within $1''$ are bound for K2 planet candidate hosts with varying background stellar densities. In addition, the absence of Gaia parallax and proper motion data suggests problems with the astrometric fit, which can be caused by relative motion of a close binary.

The right hand side of Figure~\ref{fig:Gaiacomps} shows the separation and magnitude difference of companions detected with Gaia, as well as the companions of TOI-2221 previously detected via CPM (see Section~\ref{hrcomps}). Companions identified with both Gaia and high-resolution imaging are shown in light blue, companions not recovered by Gaia are plotted in orange, and unbound companions are shown as $\times$'s. The figure highlights Gaia's inability to resolve binaries closer than $\sim 0\farcs8$, as noted by \citet{2018AJ....156..259Z}. Furthermore, nearly half of all companions (45\%) within 3\farcs5 are not detected by Gaia. The survey of M~dwarfs by \citet{2024AJ....167...56C} found an even higher rate, with nearly 60\% of companions detected with speckle imaging unresolved by Gaia. 

\begin{figure}[b]
    \centering
    \includegraphics[width=0.95\textwidth]{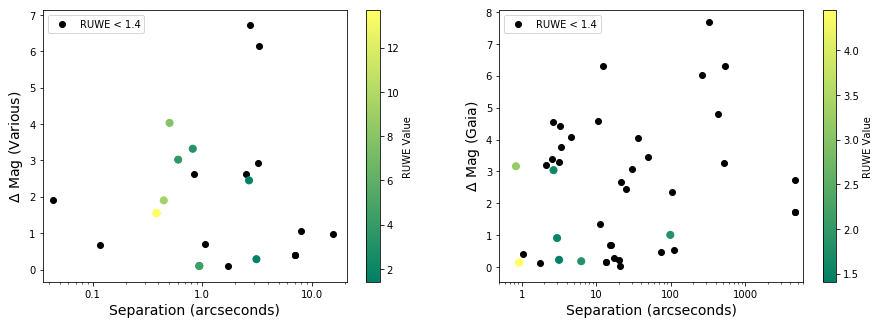}
    \caption{\textit{Left}: Magnitude difference as a function of angular separation for companions detected via high-resolution imaging, colored based on the primary star's Gaia DR3 RUWE value. Black points depict stars with RUWE values $<1.4$, indicating a high quality fit with a single star astrometric model, while colored points are scaled by their RUWE values. \textit{Right}: Magnitude difference as a function of angular separation for companions identified using Gaia DR3, including the CPM companions of TOI-2221. The points are similarly colored based on their RUWE values. Companions detected with both methods appear on both plots. Roughly half of the systems with companions detected have RUWE $>$ 1.4, making it worthwhile for preliminary multiplicity searches but not a conclusive method.}\label{fig:ruweplot}
\end{figure}

Companions below Gaia's resolution limit can be revealed through the renormalized unit weight error (RUWE) as the motion of the center of light and the center of mass differ in such cases, resulting in a poor model fit using single star astrometry \citep{2020MNRAS.496.1922B}. Single sources typically have RUWE values near one, with a value of 1.4 used as the threshold for reliable astrometric solutions. Twenty-four of the M~dwarfs in our sample have RUWE values larger than $1.4$, ten of which have one or more detected companions. High-resolution imaging is available for all but three of these systems, indicating companions are detected in about half of the stars (47\%) with elevated RUWE values. This is significantly lower than the 84\% of TOIs with RUWE~$>$~$1.4$ found to have companions in \citet{2020AJ....159...19Z}. However, 92\% of their sample was made up of solar type stars, which are more likely to host companions, and nearly one-fourth of the targets have since been confirmed as false positives \citep{2021AJ....162..192Z}, indicating contamination by close binaries. 

Of the nine companions detected within $\sim1''$ of M~dwarfs in our sample, six have RUWE $> 1.4$. All six are within 200pc and have projected separations less than 160 AU, underscoring Gaia's sensitivity to unresolved binaries is a function of the star's distance as well as the the binary separation and mass \citep{2020MNRAS.496.1922B}. This can also be seen in Figure~\ref{fig:ruweplot} where the magnitude difference as a function of separation for high-resolution and common proper motion companions is color coded by RUWE values of the host star, with higher RUWE values corresponding to nearby companions with smaller magnitude differences. 

Although many of the companions detected with high-resolution imaging but unresolved by Gaia have RUWE~$>1.4$, 33\% have values near 1.0 and would have been missed without additional imaging. This is similar to the $\sim$22\% reported by \citet{2024AJ....167...56C}. Just over half of the stars with RUWE $>1.4$ do not have a companion detected by high-resolution imaging and require further investigation to establish their multiplicity.

\section{Results}\label{sec:results}

\subsection{Stellar Properties} \label{sec:starprop}

To estimate stellar properties of the individual stars we use the TIC effective temperature reported on ExoFOP and the Modern Mean Dwarf Stellar Color and Effective Temperature Sequence\footnote{\url{https://www.pas.rochester.edu/~emamajek/EEM_dwarf_UBVIJHK_colors_Teff.txt}} based on \citet{2013ApJS..208....9P}. Effective temperatures for cool dwarfs in the TIC are calculated from Gaia $G_{\mathrm{BP}}$ and $G_{\mathrm{RP}}$ magnitudes using custom relations from \citet{2013ApJ...779..188M}, which may be contaminated by the presence of an unresolved companion \citep{2019AJ....158..138S}. As all of the companions are expected to be M~dwarfs (see below) and only six were unresolved by Gaia, we do not expect any blended photometry to significantly impact on our results.

For companions detected with AO, which have a delta magnitude in the Br-$\gamma$ or K-band, we use the effective temperature of the TOI and the Modern Mean Dwarf Sequence \citep{2013ApJS..208....9P} to determine the apparent K magnitude of the primary and use the measured delta magnitude to determine the K magnitude of the secondary. Utilizing the distance to the system, we then calculate the absolute magnitude of the secondary and use the Modern Mean Dwarf Sequence to determine the spectral type and mass. For companions without AO observations, we use the TOI effective temperature to determine the apparent I magnitude of the primary and use the 832 nm or I-band delta magnitude to determine the magnitude of the secondary. The spectral type and mass of the secondary are then determined from the Modern Mean Dwarf Sequence based on the absolute magnitude of the secondary. Estimated apparent K or I-band magnitudes for the primaries and secondaries, as well as the individual masses are shown in Table~\ref{tab:final_comp}.

\begin{figure}[b]
    \centering
    \includegraphics[width=0.4\textwidth]{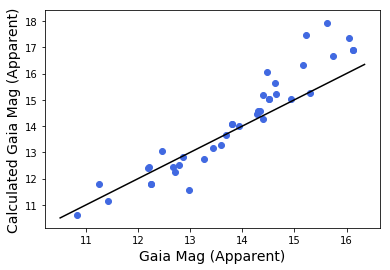}
    \caption{A comparison of archive and calculated Gaia magnitudes for common proper motion companion systems. The calculated value is based on the TIC effective temperature, which returns a characteristic magnitude independently determined using the parallax of the TOI and absolute magnitudes from the Modern Mean Dwarf Sequence \citep{2013ApJS..208....9P}. By a Gaia magnitude of 15, the values begin to diverge, with an average difference of 0.42 for $\mathrm{G}<15$ and an average difference of 1.34 for $\mathrm{G}>15$.}\label{fig:Gaiamags}
\end{figure}

Although the CPM companions identified in Gaia DR3 have magnitudes for each component, for consistency, we derive stellar parameters of the primary and secondary using the same method employed for high-resolution imaging. The TIC effective temperature is therefore used to estimate the spectral type and absolute G magnitude of the primary via the Modern Mean Dwarf Sequence, which is combined with the distance to the primary to determine the apparent G magnitude. Using the magnitude difference of the two components, we then determine the apparent G magnitude of the secondary, as well as the spectral type and estimated mass. The estimated apparent G magnitudes and the individual masses for each CPM pair are listed in Table~\ref{tab:final_comp}. Figure~\ref{fig:Gaiamags} shows that the estimated G magnitudes are generally consistent with the magnitudes measured by Gaia for brighter stars, however, the estimated magnitudes are fainter for stars with G $\gtrsim$ 15. For the nine systems identified with Gaia and high-resolution imaging, the parameters derived for the secondary components vary by one spectral subtype or less, with the spectral type determined from high-resolution imaging the same or later by 0.5. 

To provide estimates of the uncertainties in our derived stellar properties we use the uncertainty in the effective temperature to identify the range of possible masses and absolute magnitudes in the Modern Mean Dwarf Sequence and propagate them through our calculation of apparent magnitudes and stellar masses for the primary. For the companion we propagate the apparent magnitude of the primary and its uncertainty with the measured delta magnitude and its uncertainty. We use the individual uncertainties in $\Delta$m as reported on ExoFOP for all AO observations, assume 0.5~mag for all speckle observations (no uncertainties are provided for HRCam delta magnitudes), and use the median uncertainties in G-band magnitudes reported by \citet{2021A&A...649A...3R} for companions identified through CPM. The average uncertainties on the derived apparent magnitudes and stellar masses are 1.02 and 0.07, respectively, which correspond to $1 - 2$ spectral subtypes.

Four of the M~dwarf TOIs have CPM companions where the magnitude of the companion is brighter (see Table~\ref{tab:gaiacomps}), implying the TOI is the secondary component. However, the estimated stellar parameters for three of the companions show they are also M~dwarfs. We therefore keep these systems in our M~dwarf sample, but exclude TOI-468 as the estimated properties of the companion imply it is a B star.

\begin{figure}[h]
    \centering
    \includegraphics[width=0.9\textwidth]{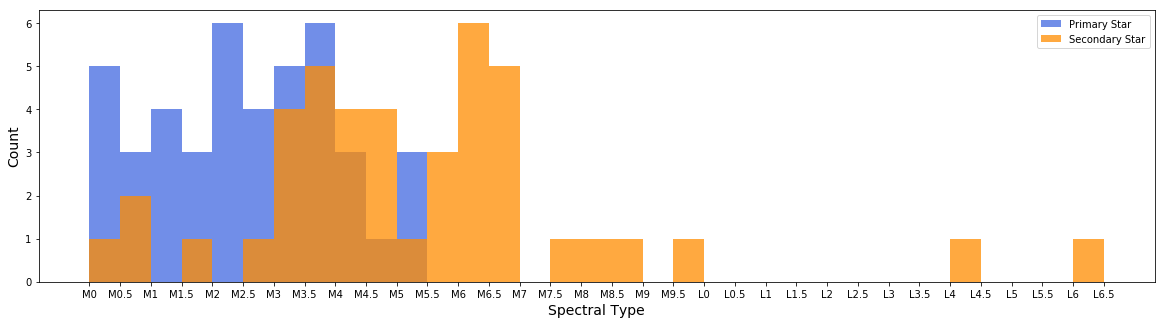}
    \caption{The distribution of estimated spectral types for all systems with stellar companions. By referencing the Modern Mean Dwarf Sequence \citep{2013ApJS..208....9P}, we use the TIC effective temperature to estimate the spectral type. The two overlapping histograms are centered around M2V-M2.5V (blue; primary stars) and M5V (orange; companions), respectively. The primary stars we detect are earlier than M5 due to the observational limits of TESS.}\label{fig:SpT} 
\end{figure}

Histograms of the estimated spectral types for the primary (blue) and secondary (orange) components are shown in Figure~\ref{fig:SpT}. The primary stars range from M0 $-$ M5, highlighting the magnitude limits of TESS. Nearly all M~dwarf spectral types are represented in the companions, consistent with the results of \citet{2019AJ....157..216W}, and appear in a Gaussian-like distribution that ranges from M0 $-$ M9 and peaks around M5. Two especially faint companions detected around TOI-4446 have magnitudes consistent with late L type dwarfs, which we do not include in our binary star analysis.

\startlongtable
\begin{longrotatetable}
\begin{deluxetable*}{lcccccccCC}
\tabletypesize{\footnotesize}
\tablewidth{0pt}
\tablecaption{Observed and derived properties of analyzed companions\label{tab:final_comp}}
\tablehead{
\colhead{TOI} & \colhead{Separation ($''$)} & \colhead{Pri.~Mag (filter)} & \colhead{Sec.~Mag (Calc.)} & \colhead{T$_{\mathrm{eff}}$ (K)} & \colhead{Distance (pc)} & \colhead{Proj.~Sep (AU)} & \colhead{$M_{\mathrm{pri}}$ ($M_{\odot}$)} & \colhead{$M_{\mathrm{comp}}$ ($M_{\odot}$)} & \colhead{RUWE}} 
\startdata 
  218 &            13.53 &       14.63 (G) &         14.79 &                3146 &        52.56 &             710.96 &  0.18 & 0.18 &   1.12 \\
  277\tablenotemark{a} &            17.25 &       12.47 (G) &         12.78 &                3748 &        64.91 &            1119.47 &  0.57 & 0.54 &   1.28 \\
  455 &             7.00 &        8.04 (I) &          8.43 &                3562 &         6.86 &              48.05 &  0.44 & 0.4 &   1.07 \\
  455 &             8.00 &        8.04 (I) &          9.09 &                3562 &         6.86 &              54.91 &  0.44 & 0.27 &   1.07 \\
  457 &             1.06 &       10.87 (I) &         11.93 &                3054 &       125.36 &             130.38 &  0.16 & 0.12 &   1.03 \\
  488 &            49.26 &       12.45 (G) &         15.92 &                3332 &        27.31 &            1345.51 &  0.27 & 0.09 &   1.30 \\
  507 &            73.46 &       14.48 (G) &         14.95 &                3338 &       109.94 &            8076.36 &  0.27 & 0.23 &   1.10 \\
  737 &             0.84 &        6.55 (K) &          9.17 &                3394 &       291.99 &             245.86 &  0.37 & 0.10 &   1.01 \\
  756 &            11.09 &       13.68 (G) &         15.02 &                3614 &        86.13 &             955.37 &  0.47 & 0.37 &   1.24 \\
  762 &             3.20 &       14.93 (G) &         18.22 &                3399 &        98.84 &             315.99 &  0.37 & 0.12 &   1.03 \\
  864 &             0.04 &        8.30 (I) &         10.22 &                3460 &        38.01 &               1.67 &  0.40 & 0.18 &   1.18 \\
 1215 &             1.73 &        7.31 (I) &          7.41 &                3751 &        34.62 &              59.76 &  0.54 & 0.54 &   1.31 \\
 1227\tablenotemark{a} &           511.83 &       11.96 (G) &         15.22 &                3050 &       100.96 &           51675.47 &  0.44 & 0.16 &   1.06 \\
 1450 &             3.25 &        6.55 (K) &          9.49 &                3407 &        22.44 &              73.01 &  0.37 & 0.09 &   1.01 \\
 1452 &             3.13 &        7.10 (K) &          7.38 &                3248 &        30.50 &              95.60 &  0.27 & 0.23 &   1.41 \\
 1634 &             2.54 &        6.18 (K) &          8.80 &                3455 &        35.07 &              89.01 &  0.40 & 0.12 &   1.23 \\
 1696 &           426.19 &       15.31 (G) &         20.11 &                3181 &        64.62 &           27540.46 &  0.23 & 0.08 &   1.12 \\
 1746 &             0.94 &        7.10 (K) &          7.20 &                3340 &        44.08 &              41.57 &  0.27 & 0.27 &   4.45 \\
 1763 &            15.11 &       13.94 (G) &         14.65 &                3520 &        88.14 &            1331.82 &  0.44 & 0.37 &   1.26 \\
 1801 &           105.05 &       10.83 (G) &         13.19 &                3815 &        30.89 &            3245.32 &  0.57 & 0.27 &   1.04 \\
 1883 &            15.66 &        11.53 (K) &        12.50 &                3477 &       117.53 &            1840.60 &  0.40 & 0.27 &   1.12 \\
 1883 &            261.14 &        14.50 (G) &          20.52 &                3477 &       117.53 &         30693.06 &       0.4 & 0.08 & 1.12 \\
 2068 &            19.99 &       12.21 (G) &         12.44 &                3700 &        53.00 &            1059.22 &  0.50 & 0.47 &   1.16 \\
 2072 &             2.99 &       12.70 (G) &         13.61 &                3546 &        39.02 &             116.62 &  0.44 & 0.37 &   1.54 \\
 2084 &            12.25 &       14.39 (G) &         20.68 &                3630 &       114.29 &            1399.48 &  0.47 & 0.09 &   1.06 \\
 2094 &            10.71 &       13.43 (G) &         18.03 &                3457 &        50.22 &             538.08 &  0.40 & 0.10 &   1.32 \\
 2205 &            2.15 &       16.13 (G) &         19.32 &                3653 &       414.95 &            892.04 &     0.50 & 0.184 & 1.05 \\
 2205 &            21.83 &       16.13 (G) &         18.79 &                3653 &       414.95 &            9059.55 &  0.50 & 0.23 &   1.05 \\
 2221 &             4680.50 &        8.04 (G) &          9.78 &                3588 &         9.71 &              45466.82 &  0.44 & 0.23 &   0.93 \\
 2221 &             4682.20 &        8.04 (G) &         10.76 &                3588 &         9.71 &              45483.34 &  0.44 & 0.16 &   0.93 \\
 2267 &             0.38 &       10.87 (I) &         12.42 &                3022 &        22.55 &               8.66 &  0.16 & 0.10 &  13.71 \\
 2293 &             4.66 &       12.86 (G) &         16.93 &                3668 &        62.75 &             292.63 &  0.50 & 0.16 &   1.23 \\
 2384 &             0.82 &      7.63 (I) &         10.96 &              3678 &       187.90 &             154.46 &  0.5 & 0.16 &   3.21 \\
 2455 &             0.51 &        5.98 (K) &         10.01 &                3553 &       100.98 &              50.99 &  0.44 & 0.09 &   7.79 \\
 2781 &             0.45 &        7.03 (I) &          8.93 &                3868 &       161.21 &              72.08 &  0.57 & 0.37 &   9.34 \\
 3397 &           536.57 &       14.29 (G) &         20.59 &                3811 &       181.78 &           97537.93 &  0.57 & 0.09 &   0.99 \\
 3494 &             0.59 &        9.78 (I) &         12.68 &                3230 &        34.07 &              20.19 &  0.23 & 0.09 &   3.72 \\
 3714 &             2.67 &       14.29 (G) &         18.85 &                3538 &       113.14 &             302.05 &  0.44 & 0.10 &   1.15 \\
 3984 &             3.27 &       14.65 (G) &         19.08 &                3422 &       108.88 &             356.11 &  0.37 & 0.09 &   1.10 \\
 4325 &            25.30 &       12.98 (G) &         15.44 &                3896 &        51.21 &            1295.73 &  0.57 & 0.27 &   1.18 \\
 4336 &             6.30 &       12.25 (G) &         12.43 &                3365 &        22.45 &             141.47 &  0.37 & 0.37 &   1.86 \\
 4336 &             98.50 &       12.25 (G) &         13.26 &                3365 &        22.45 &             2211.68 &  0.37 & 0.23 &   1.86 \\
 4349 &             2.68 &        5.15 (K) &          7.60 &                3851 &        71.99 &             193.01 &  0.57 & 0.23 &   1.57 \\
 4642\tablenotemark{a} &           110.77 &       12.73 (G) &         13.27 &                3266 &        23.56 &            2609.66 &  0.37 & 0.27 &   1.35 \\
 4668 &            20.91 &       15.16 (G) &         15.19 &                3210 &       106.00 &            2216.60 &  0.23 & 0.23 &   1.06 \\
 4858 &            36.34 &       16.05 (G) &         20.11 &                3333 &       198.49 &            7212.79 &  0.27 & 0.09 &   1.03 \\
 4991 &           165.15 &       15.75 (G) &         20.47 &                3601 &       343.08 &           56658.57 &  0.47 & 0.18 &   1.04 \\
\enddata
\tablenotetext{a}{The TOI is fainter than the detected companion and assumed to be the secondary.}
\tablecomments{Parenthetical notes in the primary magnitude column indicate what band the data are in. I = I Band, K = K Band, G = Gaia Band}
\end{deluxetable*}
\end{longrotatetable} 

 \subsection{Binary Parameters}

\begin{figure}[t!]
    \centering
    \includegraphics[width=0.95\textwidth]{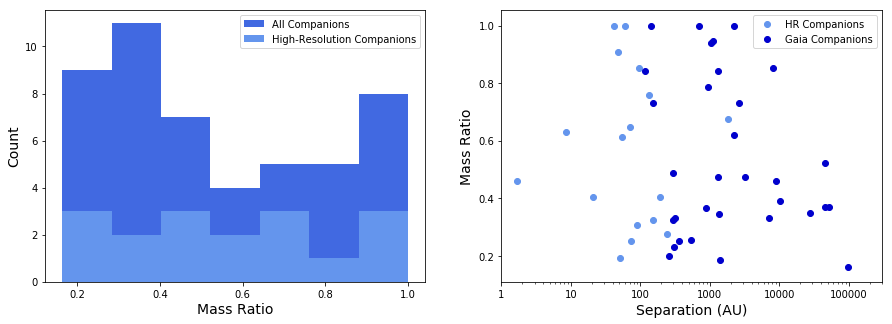}
    \caption{\textit{Left:} A histogram of mass ratios for all companions relative to the primary star, with the high-resolution companion distribution highlighted. We estimate the mass ($M_{\odot}$) by spectral type from \citet{2013ApJS..208....9P}. All included systems consist of two M~dwarfs. As such, any system where the companion was brighter than the primary (all of which are CPM pairs), we switched designation to maintain q $\leq$ 1. \textit{Right:} Mass ratio as a function of separation in au of all detected stellar companions. Between separations of 40AU to 10000AU, we find a uniform distribution of mass ratio and separation. Above a separation of 10000AU, there is a lack of companions with a mass ratio above 0.5 and an apparent slight preference for lower q with increasing separation. Below a separation of 40AU, there is a distinct drop off in the number of companions, with those present evenly split on either side of q $=$ 0.5. 
    }\label{fig:massratio}
\end{figure}

Using the individual estimated masses we evaluate the mass ratio for 47 companions relative to their primaries, assuming the primary is the brighter star. The distribution of mass ratios is shown on the left side of Figure~\ref{fig:massratio}, with companions detected via high-resolution imaging shown in light blue. The companion detection method does not correspond to a distinct physical boundary between the samples, however, assessing them separately allows for comparisons with other studies using similar techniques while also examining trends in close vs.~wide binaries. 

More than a dozen studies of M~dwarf multiplicity have been carried out in the past $\sim$ 40 years \citep[e.g.,][and references therein]{2019AJ....157..216W, 2024AJ....167..174C}, with varying sample sizes and resolution limits. The AstraLux high-resolution imaging survey of \citet{2012ApJ...754...44J} detected companions to M0 $-$ M5 primaries with separations of 1.0 - 6\farcs0 ($\sim 3 - 230$AU), similar to the companions found by high-resolution imaging within 0.04 - 4\farcs0 ($1.5 - 245$AU) in this work. They determined the binary mass ratio distribution of their sample to be more consistent with a uniform distribution than one that rises toward equal mass systems, in agreement with the mass ratio distribution for high-resolution systems seen in Figure~\ref{fig:massratio}. Furthermore, \citet{2019AJ....157..216W} surveyed all M~dwarfs within 25~pc for companions $>2''$ and found the mass ratio distribution for primaries with masses of 0.3 $-$ 0.6 M$_{\sun}$ ($\sim$ M0 $-$ M4) consistent with a flat distribution.

The mass ratio distribution for our full sample peaks at $q \sim 0.3$, consistent with wide companions to M dwarfs being skewed toward smaller masses and the wide binary fraction increasing with primary mass between $0.1 - 0.5$ M$_{\sun}$ \citep{2023ASPC..534..275O}. 
In general, the mass ratio distribution agrees with the broken power law parameterizations of \citet{2017ApJS..230...15M} and \citet{2019MNRAS.489.5822E}, and reflects the dependence on separation found in \citet{2019MNRAS.489.5822E} as wider companions favor smaller mass ratios. An apparent increase in equal mass systems may reflect the preference for equal mass binaries in mid-M primaries \citep{2023ASPC..534..275O} as well as the expected twin excess for binaries with close and intermediate separations \citep{2019MNRAS.489.5822E}. Examining the mass ratio as a function of binary separation in the right side of Figure~\ref{fig:massratio}, however, shows a deficit of like-mass pairs below separations of $\sim 40$AU, potentially reflecting the infrequency of close binaries in planet hosts or observational biases in the sample. For instance, nearby bright stars impact the photometry and astrometry of planet candidate host stars and may cause them to be rejected, biasing the sample toward faint, low mass companions \citep{2013MNRAS.428..182B}.

Although we do not account for any observation or selection biases in our sample, the mass ratio distribution of M~star exoplanet host binaries appears consistent with that of field M~dwarfs. This is in agreement with similar studies of solar-type exoplanet host stars. Transiting planet hosts with relatively close companions detected using high-resolution imaging \citep[e.g.,][]{2013MNRAS.428..182B, 2020AJ....159...19Z, 2021AJ....161..134H, 2021AJ....161..164H, 2021AJ....162...75L, 2021AJ....162..192Z} show mass ratios consistent with the roughly uniform distribution of solar-type binaries \citep{2010ApJS..190....1R, 2017ApJS..230...15M}, while companions detected around exoplanet host stars using common proper motions show a peak at $q \sim 0.3$ \citep{2022AJ....163..160B}. 

\begin{figure}[t]
    \centering
    \includegraphics[width=0.95\textwidth]{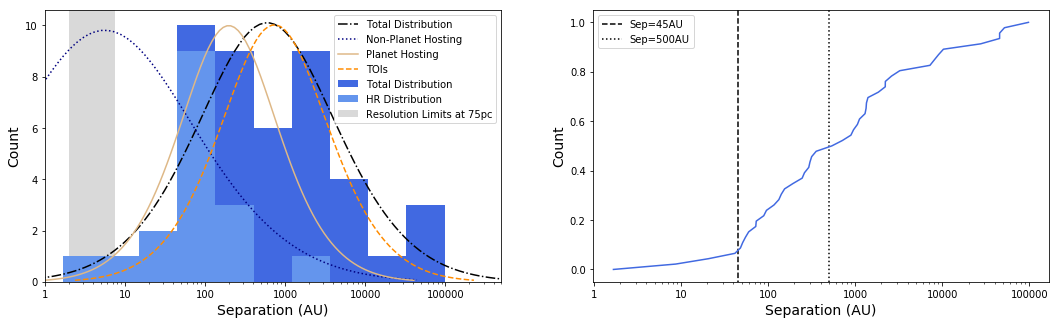}
    \caption{\textit{Left:} The projected separation distribution in au for M~dwarf TOI's with companions. The subset of companions detected with high-resolution (HR) is highlighted in light blue. We converted the observed separations of each system from arcseconds to  au using parallaxes from Gaia, then fit both the total sample (dot dash black line) and the high-resolution subset with Gaussian curves with $\sigma =$ 1.06 and $\sigma =$ 0.65 $\log$(AU), respectively. For the high-resolution sample, the peak is $\mu =$ 66~au,  whereas for the total sample, the peak is $\mu =$ 596~au. The separation distributions of non-planet hosting, planet hosting, and TOI M~stars from \citet {2024AJ....167..174C} are shown for comparison. The region shaded in gray highlights the inner angular resolution limits of the various high-resolution imaging techniques at 75pc, the mean distance for M~dwarfs in our sample. These limits demonstrate that we are sensitive to stellar companions with separations of a few au where the M-dwarf distribution of \citet {2024AJ....167..174C} peaks.   
    \textit{Right:} A normalized cumulative distribution function for the projected companion separation in au. The function highlights the deficit of companions within 45 au (indicated by the vertical dashed line). Half of the M~dwarf TOI companions are within 500  au (denoted by the vertical dotted line), reflecting the high concentration of companions between $45-500$ au as seen in the histogram on the left. The change in slope at $\sim50$~au is consistent with the semi-major axis cutoffs for close binaries with exoplanets determined by \citet{2016AJ....152....8K} and \citet{2021AJ....162..192Z}.
    }\label{fig:logsep}
\end{figure}

Next, we calculate the projected physical separation of each companion from the primary star in astronomical units using the observed angular separation and Gaia distance. The separation distribution, shown on the left in Figure~\ref{fig:logsep}, peaks at $\sim$600 au while the companions detected with high-resolution imaging (light blue) peak at 66\,AU. M~dwarf binaries in the sample of \citet{2019AJ....157..216W} show a broad peak in separation centered at 20 au, while investigations that probe closer companions find peaks at smaller separations, including $\sim 15$ au in \citet{2012ApJ...754...44J} and $\sim 6$~ au in \citet{2024AJ....167..174C}. Despite the similar angular separations probed by our high-resolution imaging and the study of \citet{2012ApJ...754...44J}, the peak in our separation distribution is $\sim4.5\times$ larger. An even more drastic difference is seen between the peak in our full distribution and that of \citet{2024AJ....167..174C}, which includes companions with separations of $0.07 - 300''$ to all M~dwarfs within 15pc. In particular, they find many more stellar companions to M~dwarfs at close separations than we do, especially with near equal magnitudes (see Figure 7 in \citealt{2024AJ....167...56C}). Although the M~dwarfs in our sample are further away, with an average distance of 75 pc and only 15 stars within 15pc, speckle imaging allows us to probe down to $\sim$2 au for two-thirds of our sample, which is below the peak in projected separation found by \citet{2024AJ....167..174C}. The inner angular resolution limits of all high-resolution imaging techniques used in this survey are highlighted in gray in Figure~\ref{fig:logsep}, which, at a distance of 75pc,  span separations of $2 - 7.5$ au. Thus, the shifted peak in our separation distribution can be attributed to fewer close companions in our sample and not our lack of sensitivity to such companions. Furthermore, \citet{2024AJ....167..174C} divided their sample by stars with and without known planets, determining statistically significant differences in the separation peaks at 198 au and 5.6 au, respectively. The distributions for the non-planet hosting and planet hosting M~dwarfs are plotted in Figure~\ref{fig:logsep}, as well as the separation distribution for M~dwarf TOIs from \citet{2022AJ....163..232C, 2024AJ....167..174C}. Our separation distribution is consistent with those of the two planet hosting distributions, but differs considerably from the separation distribution of non-planet hosting binaries.

This increase in the projected separation peak for companions orbiting exoplanet host stars relative to field stars has been observed in FGK \citep{2021AJ....161..164H, 2021AJ....162...75L} and M stars \citep{2022AJ....163..232C, 2024AJ....167..174C} and supports theoretical work that close binary companions influence planet formation. \citet{2016AJ....152....8K} and \citet{2020AJ....159...19Z,2021AJ....162..192Z} searched for stellar companions to exoplanet host stars from Kepler and TESS, respectively, and found a deficit of stellar companions within $\sim50$~au. \citet{2021MNRAS.507.3593M} compiled various surveys to determine the ratio of exoplanet host binaries to field binaries as a function of separation, which they determine increases linearly in log from $0-15$\% for $1-10$ au and $15-100$\% for $10 - 200$au. In our sample of M~dwarf TOIs 16 of the detected companions (38\%) are closer than 200au, however, we detect only five within 50au. The lack of companions within $\sim50$ au is apparent in the normalized cumulative distribution function of projected separations shown on the right side of Figure~\ref{fig:logsep}, which shows a distinct change in slope at 45 au (indicated by a vertical dashed line). 

In contrast, common proper motion studies of wide binaries find a peak in the projected separation distribution at $\sim 1000$ au \citep{2018MNRAS.480.4884E, 2020ApJS..247...66H} or larger \citep[e.g.,][]{2015AJ....150...57D} depending on the survey limits. A search for stellar companions to TESS exoplanet host stars using Gaia data by \citet{2022AJ....163..160B} found most companions located within $300 - 4000$ au and a peak near 1000au, while \citet{2021FrASS...8...16F} found a peak around 600 au for co-moving companions to exoplanet host stars within 200pc. The peak of our full sample occurs at slightly smaller separations, likely influenced by the closer separations of M~dwarf binaries and the inner angular separation limit of our study. 

The binary properties of M~dwarf exoplanet hosts are, therefore, largely consistent with field binaries in terms of mass ratio and their overall (wide-binary) separation distribution, implying close-in planet formation is possible in a variety of binary systems. The companion separation distribution for close binaries, however, is shifted to larger separations in agreement with the observed under abundance of close-in stellar companions to transiting exoplanet hosts.

\section{Discussion}\label{sec:discussion}

\subsection{Exoplanet Host Star Multiplicity}\label{sec:multiplicity}

We detected 47 companions around 42 M~dwarf TOIs using high-resolution imaging and common proper motions resulting in a multiplicity fraction of $19.4\pm2.7$\%. These findings are generally consistent with the M~dwarf multiplicity rates of $23.7\pm1.3$\% and $23.5\pm2.0$\% from \citet{2019AJ....157..216W} and \citet{2024AJ....167..174C}, respectively. Based on simulations by \citet{2024AJ....167...56C} showing speckle observations detect approximately 70\% of stellar companions to M~dwarfs and the corrected multiplicity fraction of $26.8\pm1.4$\% in \citet{2019AJ....157..216W}, we estimate our survey completeness is around $80-85$\%.

Our overall multiplicity rate also agrees with the surveys of \cite{2023AN....34430055M} and \citet{2024MNRAS.527.3183M}, which found $19.9\pm1.5$\% and $19.2\pm0.9$\% of TESS TOIs and confirmed exoplanet host stars, respectively, have companions based on Gaia astrometry and close companions in the literature. Our results are also comparable to the value of $23.2\pm1.6$\% obtained by \citet{2021FrASS...8...16F} in a similar study. Although our sample is limited to M~dwarfs, the overall binary fraction appears consistent with the wide-binary fraction seen in field stars \citep[$15 - 23\%$,][]{2010AJ....139.2566D, 2021MNRAS.507.3593M} and exoplanet host stars.

Of the 216 M~dwarfs in our final sample, 192 were observed with high-resolution imaging (89\%), resulting in a multiplicity rate of $10.0\pm2.2$\%. If we include only companions within $3\farcs0$, similar to other high-resolution imaging surveys, our multiplicity rate is $7.8\pm1.9$\%. This is approximately half that determined for solar-type exoplanet hosts by \citet[][$18.3\pm1.4$\%]{2021AJ....162..192Z} and \citet[][16\%]{2021AJ....162...75L}, which is consistent with the difference in multiplicity rates for field M~dwarf \cite[$24\%$]{2019AJ....157..216W, 2024AJ....167..174C} vs.~solar-type binaries \citep[][$46$\%]{2010ApJS..190....1R}. However, it is only 60\% of the fraction of companions detected within 3\farcs5 of nearby ($\lesssim100$ pc) M dwarfs by \citet[][13.4\%]{2024AJ....167...56C}.

These results show that short-period planet formation is a routine occurrence in binary stars, as our multiplicity rate is generally consistent with M~dwarfs without known planets, as well as studies of CPM companions with and without known planets. However, the deficit of close companions observed in exoplanet host systems is clearly seen in our sample of M~dwarf TOIs as we detect significantly fewer close companions than \citet{2024AJ....167...56C, 2024AJ....167..174C} despite being sensitive to companions within a few au for the majority of our sample.

\subsection{Planet Properties} 

To investigate trends in the properties of close-in transiting exoplanets in multiple star systems we retrieve the TOI planet parameters from ExoFOP\footnote{As of 8 February 2024.} for the 260 exoplanets in our sample. We use the planet radius and orbital period determined by the TESS project for all TOIs except the confirmed planets in TOI-455 \citep{2022AJ....163..168W} and TOI-2221 \citep{2020Natur.582..497P, 2021A&A...649A.177M}, where we use parameters from the cited publications as the ExoFOP data are incomplete. For host stars with nearby companions the additional flux will dilute the observed transit depth resulting in underestimated planetary radii \citep{2015ApJ...805...16C} and overestimated densities \citep{2017AJ....154...66F}. Planet parameters provided by the TESS team are corrected for contamination from known neighboring stars based on the TIC \citep{2021ApJS..254...39G}. However, six of the high-resolution imaging companions are not listed in the TIC and/or the TOIs do not have contamination ratios provided. We compute radii correction factors for these planets based on \citet{2015ApJ...805...16C}
and provide updated radii in Table~\ref{tab:radii_cor} assuming the planet orbits the brighter, primary star. Attempts to determine which star the transiting planet orbits in a multi-star system show the primary star is the exoplanet host in most cases, with at least $60 - 70$\% of transiting exoplanet host stars orbiting the primary \citep{2018AJ....156..209P, 2022AJ....164...56L}. Our corrected radii are 12\% larger on average, though if the planets orbit the secondaries they could be significantly larger, underscoring the impact stellar companions have on the occurrence rates of small planets.

\begin{figure}[]
    \centering
    \includegraphics[width=0.9\textwidth]{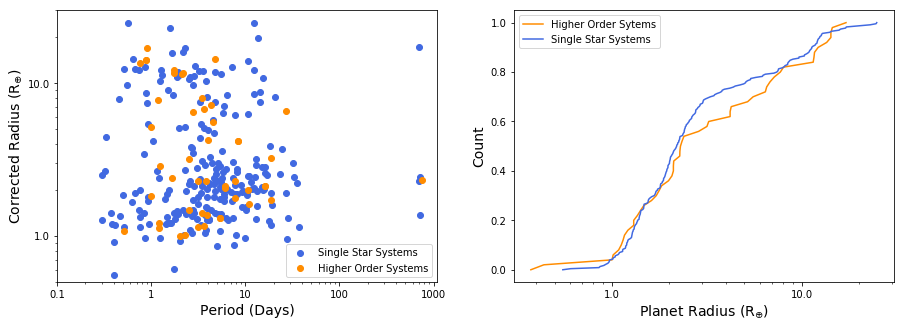}
    \caption{\textit{Left:} A comparison of the radii of candidate and confirmed transiting planets around single-star and higher order systems as a function of the planet’s period. Here we use the corrected radii shown in Table~\ref{tab:radii_cor} for the higher order systems. The distribution of planets in multi-star systems appears similar to that of planets around single stars, although the observational bias against detecting small planets in multiple star system is noticeable at P $\gtrsim 4$ days.
    \textit{Right:} Normalized cumulative distribution function for planet radii in single and multiple star systems. There is no statistically significant difference between the two samples.
    }\label{fig:pl_rad}
\end{figure}

\startlongtable
\movetabledown=10mm
\begin{deluxetable*}{lccc}
\tabletypesize{\footnotesize}
\tablewidth{0pt}
\tablecaption{Corrected planetary radii for TOIs with previously unresolved companions\label{tab:radii_cor}}
\tablehead{
\colhead{TOI} & \colhead{Uncorrected Radius ($R_{\earth}$)} & \colhead{Correction Factor} & \colhead{Corrected Radius ($R_{\earth}$)}} 
\startdata 
864.01 &                          1.005 &             1.082 &                        1.087 \\
 1215.01 &                          0.879 &             1.383 &                        1.216 \\
 2267.01 &                          1.273 &             1.114 &                        1.418 \\
 2267.02 &                          0.901 &             1.114 &                        1.003 \\
 2267.03 &                          0.913 &             1.114 &                        1.017 \\
 2455.01 &                         14.327 &             1.012 &                       14.501 \\
 2781.01 &                          5.942 &             1.083 &                        6.438 \\
 3494.01 &                          2.217 &             1.034 &                        2.292 \\
\enddata
\end{deluxetable*}

The planet radius as a function of orbital period is shown in Figure~\ref{fig:pl_rad} for all single and multiple star systems in our sample. As expected for M~dwarfs the majority of the planets are small \citep[e.g.,][]{2015ApJ...814..130M,  2020AJ....159..211C, 2021A&A...653A.114S}, although there are planets as large as 17~R$_\earth$. Overall, the distribution of multi-star systems mirrors that of close-in planets orbiting single M~dwarfs, reflecting the prevalence of small planets less than 4~R$_\earth$, the sparsely populated Neptune parameter space, and the small population of giant planets seen in such systems \citep{2023MNRAS.521.3663B,2023AJ....166...44P}. The observational bias against detecting Earth-sized transiting planets in binary stars \citep[e.g.,][]{2021AJ....162...75L, 2021AJ....162..192Z}, which increases with planetary orbital period (fewer transits, lower signal-to-noise), becomes apparent for small planets with periods greater than $\sim4$ days. 

\begin{figure}[b!]
    \centering
    \includegraphics[width=0.9\textwidth]{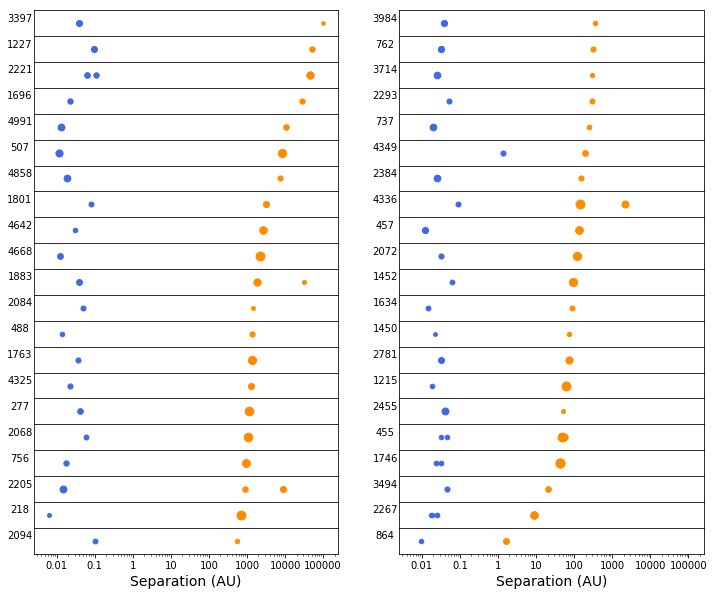}
    \caption{An examination of the projected separation for all components in higher order stellar systems. The primary star is taken to be at a separation of 0 and is not displayed. Orange dots represent stellar companions and are scaled according to the mass ratio, while blue dots represent confirmed and candidate transiting planets and are independently scaled using a $M\propto R^{3}$ mass estimation. There does not appear to be any obvious relation between the separation of the planet and the separation of the stellar companion, though we note slightly more close-in, giant planets (58\%) in systems with stellar projected separations $<$~450  au where the columns split.
    }\label{fig:pl_sep}
\end{figure}

At first glance, the planets in multiple star systems appear to be larger, with an average radius of  4.7~R$_\earth$ compared to an average of  3.4~R$_\earth$ in single stars, in alignment with previous studies showing higher-mass planets occur more often in multiple star systems  \citep[e.g.,][]{2021FrASS...8...16F, 2024MNRAS.527.3183M}. In particular, \citet{2024MNRAS.527.3183M} found that the median mass for confirmed planets in multi-star systems was seven times larger due to single stars hosting more super-Earths while Jovian planets were more common in multiple star systems.
However, a two-sample Kolmogorow–Smirnow (K-S) test of the planet radii for single vs.~multi star systems in our sample returns a p-value of 0.47 implying they are likely drawn from the same distribution. The normalized cumulative distribution function for planet radii in single and multiple star systems is shown in Figure~\ref{fig:pl_rad}, with the largest difference between the two samples (0.13) occurring near $3$~R$_\earth$. When comparing the radii of TOIs around solar-type stars, \citet{2021AJ....162...75L} found  the distributions for planets in single and multi-star systems to be statistically different (p-value of 0.001), which they attributed to the observational bias against detecting Earth-sized planets in binaries. While we see fewer small planets with  P~$\gtrsim 4$ days in multi-star systems, the bias is less severe than for solar-type stars in \citet{2021AJ....162...75L}, and may explain why we find no statistical difference between planets in single and multiple star systems. In comparison with the large compilations of confirmed planets that find more massive planets in binaries, the difference is likely because the majority of planets orbiting M~dwarfs are small, which appear to be minimally affected by the presence of a companion \citep{2021FrASS...8...16F}. Furthermore, there is evidence planets are less impacted by low mass stellar companions \citep{2024AJ....167...89Z}. However, the sensitivities and selection biases inherent in different planet detection methods (e.g., transit vs.~radial velocity surveys) can make it difficult to compare planetary trends without a thorough analysis of such effects \citep[see e.g.,][]{2021MNRAS.507.3593M}, which is beyond the scope of this work.

\begin{figure}[t]
    \centering
    \includegraphics[width=0.45\textwidth]{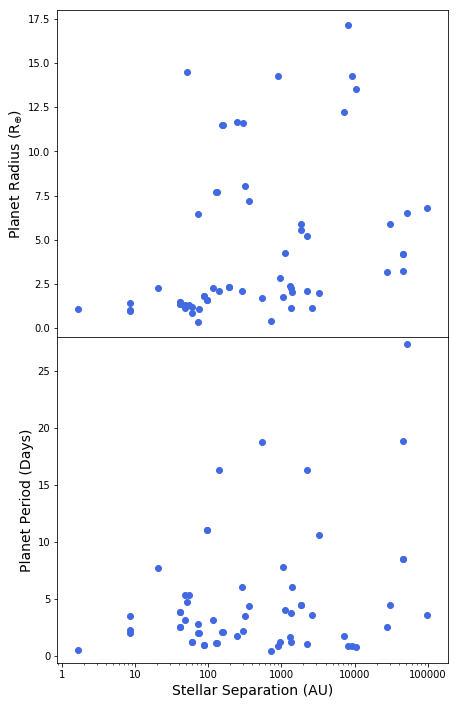}
    \caption{An examination of planetary properties (top: radius in $R_{\oplus}$; bottom: period in days) as a function of projected separation for the stellar companion. For systems with multiple stellar companions, all planets and planet candidates are compared against the primary and secondary stars. Higher order stellar companions are compared with the first planet or planet candidate of the system. A near-linear relation can be seen in the bottom plot where the range of planet periods increases with stellar separation.
    }\label{fig:pl_binsep}
\end{figure}

Figure~\ref{fig:pl_sep} shows the separations of all planetary and stellar companions around the 42 multi-star systems in our sample. All but one of the planets are within $\sim0.1$ au of the host star, consistent with the parameter space where \citet{2021FrASS...8...16F} found planet properties most impacted by stellar companions. However, there do not appear to be any significant trends between the separations of the planets and stellar companions. 
An enhancement of massive ($>$~0.1~M$_\mathrm{J}$) short-period planets have been noted in several studies \citep[e.g.,][]{2021FrASS...8...16F, 2021MNRAS.507.3593M, 2021AJ....162..272S} and attributed to the presence of a stellar companion within a few hundreds of au. Such companions are predicted to enhance planet migration and potentially allow a planet to accrete additional material as it moves inward \citep{2021AJ....162..272S} or trigger gravitational fragmentation that can lead to the formation of giant planets in otherwise stable disks \citep{2022MNRAS.511..457C}. While we do not see an obvious excess of such systems around the M~dwarfs in our sample, we note a slightly larger fraction of giant planets (58\%) on the right side of Figure~\ref{fig:pl_sep} where the stellar companions are within $<$~450 au of the host star. 

We also examine the planet radius and planet period as a function of stellar separation in Figure~\ref{fig:pl_binsep}. There is a clear lack of $\gtrsim 3$~R$_\earth$ planets in systems that have a stellar companion within $\sim50$AU, further emphasizing that close-in stellar companions inhibit planet formation, but no other obvious trends.
While we see no difference in the planetary orbital periods in single vs.~multiple star systems (K-S test p-value = 0.45), there appears to be a correlation between the binary separation and range of planetary orbital periods in multi-star systems. The bottom plot in Figure~\ref{fig:pl_binsep} shows the planet period as a function of the separation of the closest stellar companion and an increase in long period planets in wider binaries.

M~dwarfs are also known for hosting multi-planet systems, with an average of 2.5 small planets per star  \citep[1-4~R$_\earth$;][]{2015ApJ...807...45D}. However, low-mass, tightly-packed multi-planet systems are less commonly found in multiple star systems \citep{2021FrASS...8...16F}, especially in close binaries \citep{2021AJ....162..272S}. In our sample 14\% of the single star TOIs have two or more transiting planets, while a smaller fraction (9\%) of the TOIs with companions have multiple planets. Intriguingly three of the four multi-planet systems have companions detected within 50~ au (see Figure~\ref{fig:pl_sep}). While the stellar companions may be wide pairs that appear closer due to their projected separation, the fact that three of the four systems have companions at distances that are expected to be inhospitable for planets seems significant. The existence of these systems may indicate the formation of close-in planets in binary stars depends on properties of the host star (e.g.,~mass, luminosity, metallicity) or the observational bias against detecting Earth-sized planets has prevented the discovery of such planets in other systems.

\section{Conclusions}

In order to examine the demographics of M~dwarf binaries with transiting planets, we investigate a total of 308 M~dwarf TOI systems for stellar companions. After removing false positives, probable giants, and other systems lacking key information for analysis, we keep 216 of those systems. Using high resolution imaging - namely speckle interferometry and adaptive optics - and Gaia astrometry, we find a total of 47 companions in 42 systems, making our final multiplicity rate $19.4\pm2.7$\%.

As a part of our companion search, we used Gaia's RUWE value in order to prioritize targets for observation and determine how often an elevated RUWE value indicated the presence of a companion. Of those systems with a RUWE value above 1.4, which is considered the cutoff for a good fit from Gaia's pipeline, nearly 50\% returned with stellar companion detections. However, one-third of stars with companions detected within $1''$ do not have elevated RUWE values. As such, we conclude that the RUWE is useful as an initial indicator for systems of interest, but it should be supplemented with other methods for comprehensive companion searches. 

The multiplicity rate of our sample is consistent with that of field M~dwarfs, implying that planet formation is a common occurrence in binary systems. We do find, however, that the presence of a stellar companion can impact the formation and evolution of close-in planets. While field M~dwarfs have average projected separations anywhere from 5 to 20 au, our distribution of companion separations peaks at approximately 600 AU. Subsequently, there are fewer close-in companions in our sample than for field stars. We detect only five companions within 50 au, which is consistent with the findings of other works analyzing 
the multiplicity of transiting exoplanet host stars 
and suggests that the cause is physical and not an unaccounted for observational bias. 

As the presence of a stellar companion will dilute the signal of a transiting planet, we derived correction factors and updated radii for six planets that had no accounting for companion contamination in the TIC. Using these updated radii we compared the periods and radii for planets in single vs.~multiple star systems, finding no statistical difference between the two samples. For multi-star systems where the companion is within 50 au, however, we only see planets smaller than $\sim3$ R$_\earth$. We also note that despite their being very rare, we find three systems in our sample with multiple planets and multiple stars where the stars have a projected separation within 50 AU. Systems with stellar companions so close in are thought to be inhospitable to planets, so these systems could be of particular interest for further study. 

This paper analyzes a number of M~dwarf primaries ranging in spectral type from M0$-$M5. The criteria for forming our sample includes the coolest stellar objects available on the ExoFOP, so the absence of cooler M~dwarfs is still an observational issue that requires additional investigation. As late M~dwarfs are the primary star in approximately half of the multiple star systems containing late M~dwarfs \citep{2019AJ....158..152W}, there are a significant number of systems in need of further characterization. There are efforts beginning to look at this subset of stars, such as \citet{2023AJ....165..251T} and \citet{2023AJ....165..265M}, but their results lie beyond the scope of this paper. As such, it is difficult to conclude anything about the demographics or planetary occurrence rates of mid to late type M~dwarfs. As missions continue to expand the number of planet candidates and confirmed planets, and methods for observing fainter stars improve, we hope to see a sample of these mid to late M~dwarfs develop for further analysis of the occurrence rates of planets in low mass binaries.

\section*{Acknowledgements}
We thank the anonymous referee for comments that improved the quality of the final manuscript. The authors are also grateful to members of the TESS Follow-up Observing Program who have provided data to the community on ExoFOP, especially, Charles Beichman, Mark Everett, Colin Littlefield, Boris Safonov, and Andrei Tokovinin.

This research has made use of the Exoplanet Follow-up
Observation Program \citep{https://doi.org/10.26134/exofop5} website and NASA Exoplanet Archive, which are operated by the California Institute of Technology, under contract with the National Aeronautics and Space Administration under the Exoplanet Exploration Program. This work has also made use of data from the European Space Agency (ESA) mission Gaia (\url{https://www.cosmos.esa.int/gaia}), processed by the Gaia Data Processing and Analysis Consortium (DPAC, \url{https://www.cosmos.esa.int/web/gaia/dpac/consortium}). Funding for the DPAC has been provided by national institutions, in particular the institutions participating in the Gaia Multilateral Agreement. In addition, this research has made use of the Washington Double Star Catalog maintained at the U.S. Naval Observatory.

This research was carried out in part at the Jet Propulsion Laboratory, California Institute of Technology, under a contract with the National Aeronautics and Space Administration (80NM0018D0004).

Some of the observations in the paper made use of the High-Resolution Imaging instruments ‘Alopeke and Zorro, which were funded by the NASA Exoplanet Exploration Program and built at the NASA Ames Research Center by Steve B. Howell, Nic Scott, Elliott P. Horch, and Emmett Quigley. ‘Alopeke and Zorro are mounted on the North and South telescopes of the international Gemini Observatory, a program of NSF NOIRLab, which is managed by the Association of Universities for Research in Astronomy (AURA) under a cooperative agreement with the U.S. National Science Foundation. on behalf of the Gemini partnership: the U.S. National Science Foundation (United States), National Research Council (Canada), Agencia Nacional de Investigación y Desarrollo (Chile), Ministerio de Ciencia, Tecnología e Innovación (Argentina), Ministério da Ciência, Tecnologia, Inovações e Comunicações (Brazil), and Korea Astronomy and Space Science Institute (Republic of Korea).

This work was enabled by observations made from the Gemini North telescope, located within the Maunakea Science Reserve and adjacent to the summit of Maunakea. We are grateful for the privilege of observing the Universe from a place that is unique in both its astronomical quality and its cultural significance. 

Data presented were also obtained by the NESSI High-Resolution Imaging instrument operated at the WIYN Observatory by NSF NOIRLab, under the NN-EXPLORE partnership of the National Aeronautics and Space Administration and the U.S. National Science Foundation. WIYN is a joint facility of the University of Wisconsin–Madison, Indiana University, NSF’s NOIRLab, the Pennsylvania State University, Purdue University, University of California, Irvine, and the University of Missouri. The authors are honored to be permitted to conduct astronomical research on Iolkam Du’ag (Kitt Peak), a mountain with particular significance to the Tohono O’odham.

\vspace{5mm}
\facilities{Gemini:Gillett (`Alopeke), Gemini:South (Zorro), WIYN (NESSI), Hale (PHARO), Keck:II (NIRC2), ExoFOP}

\software{SciPy \citep{2020NatMe..17..261V}, NumPy \citep{2020Natur.585..357H}}

\vspace{1cm}
\appendix

\section{Additional companion detections}

As noted in Section 3.1, 87 TOIs were removed from our sample as they have been classified as false positives (FP), ambiguous planet candidates (APC), or are likely giant stars. High-resolution observations of M~dwarf TOIs conducted by our team and collaborators where a companion was detected but the TOI was excluded from our sample are reported in Table~\ref{tab:FPobs}. Additional details, including the reduced data and contrast curves, for these observations are available on the ExoFOP website.

\startlongtable
\movetabledown=10mm
\begin{deluxetable*}{lcccccc}
\tabletypesize{\footnotesize}
\tablewidth{0pt}
\tablecaption{Companions detected around FP, APC, or giant star TOIs\label{tab:initial_obs}}
\tablehead{
\colhead{TOI} & \colhead{Instrument} & \colhead{Filter} & \colhead{Separation (")} & \colhead{Position Angle ($\degr$)} & \colhead{Delta Magnitude} & \colhead{Date}} 
\startdata 
212 &                                          Zorro  &                             832  &    0.23  &   184.8  &            1.05  &                            2022-07-29  \\
212 &                                          Zorro  &                             832  &   0.924&   292.4 &             1.68  &                            2022-07-29 \\
   &                                          Zorro  &                             832  &   0.884  &   277.9  &             2.92  &                            2022-07-29  \\
224  &                                    Zorro  &                   562  &               0.1  &          224.8  &                 2.07  &                             2020-11-26  \\
        &                                    Zorro  &                   832  &              0.099  &           226.6  &                 1.0  &                            2020-11-26  \\
482 &                         NESSI  &                 832  &                           0.398  &                    267.3 &                         2.43  &                 2019-10-11 \\
   &                         'Alopeke  &                 832  &                           0.388  &                    261.6   &                         2.9  &                2022-02-11  \\
  531 &                                'Alopeke  &             832  &             0.603  &          299.5  &            2.17  &                            2022-02-16  \\
  543 &                                'Alopeke  &             832  &             0.186  &          151.9  &            4.6  &                            2022-02-11  \\
  573 &                                NIRC2  &             Br-$\gamma$  &             1.716  &          124  &            0.9162  &                            2019-05-12  \\
  &                             'Alopeke  &             562  &             1.419  &          227.6  &            1.3  &                            2020-02-15  \\
  &                             'Alopeke  &             832  &             1.519  &          234.3  &            1.25  &                            2020-02-15  \\
  749 &                                          NIRC2  &                  Kp &             1.324 &                         70 &       1.9281 &                                        2019-06-25 \\
   &                                          NIRC2  &                  J  &             1.321  &                         70  &       1.9364  &                                        2019-06-25  \\
   &                                          NIRC2  &                   H  &             1.322  &                         70  &      1.9345  &                                        2019-06-25  \\
  1234 &                                          PHARO  &                  Kcont &             2.137 &                         122 &       1.666 &                                        2023-08-10 \\
   &                                          PHARO  &                  Hcont  &             2.374  &                         106.64  &       2.181  &                                        2023-08-10  \\
 1256 &                                PHARO  &         Br-$\gamma$  &                           1.872 &                        140 &                3.493 &                            2021-11-11 \\
     &                                PHARO  &         Hcont  &                           1.872  &                        140  &                3.621  &                             2021-11-11  \\
 1639 &    'Alopeke  &    832  &                           1.267  &                      351.5  &                         5.02  &    2021-10-22 \\
 2451 &                              'Alopeke  &             832  &            0.209  &         291.6   &               3.81  &                            2022-02-12  \\
 3029 &                                          Zorro  &                             832  &                          0.072  &                       46.7  &                          3.31  &                                        2022-05-23  \\
 3528 &                                          PHARO  &                Br-$\gamma$  &                            1.57 &                      180.5 &                 0.197 &                           2021-08-08 \\
      &                                          PHARO  &                Hcont  &                            1.57  &                      180.5  &                 0.187  &                          2021-08-08 \\
 3583 &                                          NIRC2  &                             K  &                           0.589  &                      156.3  &                        4.054  &                                        2021-08-28  \\
 4261 &                                   Zorro  &                          832  &                   0.128  &                234.7  &                    6.06  &                            2022-03-21  \\
 5031 &                                   Zorro  &                          562 &                   0.648 &                227.3  &                    2.35  &                            2022-03-19  \\
     &                                   Zorro  &                          832  &                   0.655  &                226.6   &                    2.2  &                            2022-03-19  \\
\enddata
\end{deluxetable*}
 \label{tab:FPobs}

\vspace{1.5cm}
%\bibliography{new.ms}{}
\bibliography{Mstars}{}
\bibliographystyle{aasjournal}

\end{document}